\documentclass[journal]{IEEEtran}
\usepackage{ifpdf}

\usepackage{cite}

\ifCLASSINFOpdf
\usepackage[pdftex]{graphicx}
	\DeclareGraphicsExtensions{.pdf}
\else
  \usepackage[dvips]{graphicx}
  \DeclareGraphicsExtensions{.eps}
\fi

\usepackage[cmex10]{amsmath}
\usepackage{amssymb}
\usepackage{wasysym}
\usepackage{mathrsfs}
\DeclareMathAlphabet{\mathpzc}{OT1}{pzc}{m}{it}
\usepackage{algorithmic}

\usepackage{array}

\usepackage[tight,footnotesize]{subfigure}

\usepackage{stfloats}
\usepackage{url}

\usepackage{multirow}
\usepackage{algorithm}
\usepackage{color}

\newtheorem{defn} {Definition}
\newtheorem{te}{Theorem}

\newtheorem{cor}{Corollary}
\newtheorem{ex}{Example}
\newtheorem{prop}{Proposition}
\newtheorem{conj}{Conjecture}

\begin{document}

%
\title{LDPC Codes from Latin Squares Free of Small Trapping Sets}
%
%
%

\author{Dung~Viet~Nguyen, Shashi~Kiran~Chilappagari,~\IEEEmembership{Member,~IEEE},
				Michael~W.~Marcellin,~\IEEEmembership{Fellow,~IEEE},
				and~Bane~Vasi$\acute{\mathrm{c}}$,~\IEEEmembership{Senior~Member,~IEEE}
\thanks{D. V. Nguyen, M. W. Marcellin, and B. Vasi$\acute{\mathrm{c}}$ are with the Department
of Electrical and Computer Engineering, University of Arizona, Tucson,
AZ, 85719 USA (e-mail: \{nguyendv,marcellin,vasic\}@ece.arizona.edu).}
\thanks{S. K. Chilappagari was with the Department
of Electrical and Computer Engineering, University of Arizona, Tucson,
AZ, 85719 USA. He is now with Marvell Semiconductor Inc., Santa Clara, CA 95054 USA (email: shashickiran@gmail.com).}
\thanks{Manuscript received \today. This work is funded by NSF under the grants IHCS-0725403, CCF-0634969, CCF-0830245. The material in this paper is to be presented in part at the Information Theory Workshop (ITW2010). The work of S. K. Chilappagari was performed when he was with the Department of Electrical and Computer Engineering, University of Arizona, Tucson.}}

%
%

\markboth{Submitted to IEEE Transactions on Information Theory - August 2010.}%
{Nguyen \MakeLowercase{\textit{et al.}}: LDPC Codes from Latin Squares Free of Small Trapping Sets}
%



\maketitle

\begin{abstract}
This paper is concerned with the construction of low-density parity-check (LDPC) codes with low error floors. Two main contributions are made. First, a new class of structured LDPC codes is introduced. The parity check matrices of these codes are arrays of permutation matrices which are obtained from Latin squares and form a finite field under some matrix operations. Second, a method to construct LDPC codes with low error floors on the binary symmetric channel (BSC) is presented. Codes are constructed so that their Tanner graphs are free of certain small trapping sets. These trapping sets are selected from the Trapping Set Ontology for the Gallager A/B decoder. They are selected based on their relative harmfulness for a given decoding algorithm. We evaluate the relative harmfulness of different trapping sets for the sum product algorithm (SPA) by using the topological relations among them and by analyzing the decoding failures on one trapping set in the presence or absence of other trapping sets.
\end{abstract}


%
\IEEEpeerreviewmaketitle

\ifCLASSOPTIONcaptionsoff
  \newpage
\fi


\section{Introduction}
\IEEEPARstart{D}ESPITE the fact that numerous results on construction of LDPC codes \cite{ldpcBook_gallager} have been published in the past few years, this research topic remains contemporary in the field of coding theory. Researchers have focused on two main problems: (i) deriving new classes of structured codes and (ii) constructing codes with low error floor performance. 

To be efficiently encodable and decodable, the parity check matrix of an LDPC code must be structured (hence the term structured code). The construction of structured LDPC codes relies on algebraic or combinatorial objects. In many cases, the parity check matrix of a structured LDPC code can be represented as an array of permutation matrices. If the permutation matrices are circulant permutation matrices then the code is quasi-cylic (QC). Most researchers have focused on QC codes as these codes result in low encoding and decoding complexity. The encoding of these codes can be efficiently implemented using shift registers with linear complexity \cite{effEncoding}, while the decoding can be parallelized by exploiting the block structure of the parity check matrices \cite{09VLSI_2,09VLSI_3}. 

It is well-known that in order to achieve a reasonably good performance under iterative message passing decoding algorithms, the Tanner graph of an LDPC code must not contain cycles of length four. Numerous methods to form a parity check matrix such that its corresponding Tanner graph does not contain four cycles have been proposed. These methods ensure that any two rows (columns) of a parity check matrix have 1's in at most one common position. This constraint on parity check matrices is referred to in \cite{qcFiniteField_lanLin} as the \textit{row-column (RC) constraint}.

Algebraic methods of constructing QC LDPC codes usually exploit a one to one correspondence between an element of an algebraic structure, such as a group or a Galois field, and a circulant. This one to one correspondence translates the problem of constructing a parity check matrix into the problem of constructing a matrix of elements from the algebraic structure. The RC constraint is converted to a simpler constraint on the second matrix. Notable work on algebraic constructions of LDPC codes includes (but is not limited to) \cite{tannerCode_tanner,reedSol_ivana,nearShanonCode_ivana,arrayCode_fan,qcFiniteField_lanLin} with methods in \cite{qcFiniteField_lanLin} and \cite{arrayCode_fan} being the most relevant to the structured codes proposed in this paper. 

Combinatorial constructions of LDPC codes evolved from balanced incomplete block designs (BIBDs) \cite{designTheoryBook}. In these constructions, a parity check matrix is obtained from a point-block incidence matrix of a BIBD: points represent parity-check equations while blocks represent bits of a linear block code. The RC constraint is satisfied by setting the parameters of the BIBD so that no two blocks contain the same pair of points. The first class of combinatorially constructed LDPC codes was introduced by Kou, Lin and Fossorier in \cite{finiteGeometryLDPC_kouLin}. These codes are closely related to finite-geometry codes, a well studied class of codes which is used in conjunction with one-step or multiple step majority logic decoding. Other combinatorial methods of constructing LDPC codes were studied in great detail and summarized by Vasic and Milenkovic in \cite{combinatorialLDPC_vasic}.

In this paper, we give a new class of structured LDPC codes. The parity check matrices of these codes are arrays of permutation matrices which are obtained from Latin squares. These $q\times q$ matrices form a Galois field GF($q$) under some matrix operations (introduced later in this paper). Hence, our codes are different from the codes proposed by Lan \textit{et al.}\cite{qcFiniteField_lanLin}, which utilize a one to one correspondence between a $(q-1)\times (q-1)$ circulant permutation matrix and an element of the multiplicative group of GF($q$). The new class of codes contains array LDPC codes \cite{arrayCode_fan} when $q$ is a prime, but includes higher rate codes than shortened array LDPC codes \cite{highRateG8_vasic,shortenedArrayCode_milenkovic}, when the Tanner graphs are required to satisfy certain constraints. The description of the new class is not only concise and general but also makes the RC constraint trivial to satisfy. Above all, our permutation matrices are more general than circulants as the circulant property for our codes holds on indices understood as elements of GF($q$). More specifically, the permutation matrix corresponding to $\alpha^t \in \mathrm{GF}(q)$ sends the indices $(0,1,\alpha,\ldots,\alpha^{q-2})$ to $(0+\alpha^t,1+\alpha^t,\alpha+\alpha^t,\ldots,\alpha^{q-2}+\alpha^t)$. This new class of codes serves as a basis for a method of constructing codes with low error floor performance, which we shall now explain. 

By now, it is well established that the error floor phenomenon, an abrupt degradation in the error rate performance of LDPC codes in the high signal-to-noise-ratio (SNR) region, is due to the presence of certain structures in the Tanner graph that lead to decoder failures \cite{errorFloor_richarson}. For iterative decoding, these structures are known as trapping sets (see \cite{cover_milos} for a list of references).

To construct LDPC codes with provably low error floors, it is essential to understand the failure mechanism of the decoders in the high SNR region as well as to fully characterize trapping sets. These prerequisites had been met for decoders on the binary erasure channel (BEC), in which case trapping sets are known under the notion of stopping sets \cite{bec_di}. For the BEC, the definition of stopping sets is fully combinatorial and the code construction strategy is simply to maximize the size of the smallest stopping set. Such a level of understanding has not been gained for other channels of interest.

On other channels, such as the BSC or the additive white Gaussian noise channel (AWGNC), knowledge on trapping sets is far from complete due to the complex nature of iterative decoding algorithms, such as the SPA. As a result, code performance is typically improved by increasing the girth of the Tanner graph \cite{shortenedArrayCode_milenkovic,highGirthQC_yedidia,qcGirth14_koreanGuy}. The basis for these approaches is mostly constituted in two facts. First, a linear increase in the girth results in an exponential increase of the minimum distance if the code has column weight $d_v\geq 3$ \cite{recursiveApproach_tanner}. Second, trapping sets containing shortest cycles in the Tanner graph are eliminated when the girth is increased. In addition, several recent results can be used to justify the construction of a code with a large girth: the error correction capability under the bit flipping algorithms was shown to grow exponentially with the girth for codes with column weight $d_v\geq 5$ \cite{guaranteedError_shashic}; the minimum pseudo-codeword weight on the BSC for linear program decoding was also shown to increase exponentially with the girth \cite{pseudoWeight_kelley}. It is worth noting here that the minimum stopping set size also grows exponentially with the girth for codes with column weight $d_v\geq 3$ \cite{stoppingSetGirth_alon}. 

Nevertheless, for finite length codes, large girth comes with large penalty in code rate. In most cases, at a desirable code rate, the girth can not be made large enough for the Tanner graph to be free of the most harmful trapping sets that mainly contribute to decoding failures in the error floor region. These trapping sets dictate the size of the smallest error patterns uncorrectable by the decoder and hence also dictate the slope of the frame error rate (FER) curve \cite{cover_milos}. To preserve the rate while lowering error floor, a code must be optimized not by simply increasing the girth but rather by more surgically avoiding the most harmful trapping sets.

In this paper, LDPC codes are constructed so that they are free of small harmful trapping sets. We focus our attention on regular column-weight-three codes as these codes allow very low decoding complexity but exhibit very high error floor if they are not designed properly. A key element in the construction of a code free of trapping sets is the choice of forbidden subgraphs in the Tanner graph, since this choice greatly affects the error performance as well as the code rate. This choice is well determined if the Gallager A/B algorithm is used on the BSC since the necessary and sufficient conditions for a code to guarantee the correction of a given number of errors are known \cite{clw3paper,col3Part2}. However, for the SPA on the BSC and on the AWGNC, the choice of forbidden subgraphs is not clear due to the lack of a combinatorial characterization of trapping sets for these channels. In a series of papers \cite{05SCCV1,04CCSV,04CCSV2} we used the notion of \textit{instantons} to predict the error floors as well study the phenomenon from a statistical mechanics perspective. In \cite{instantonPaper_shashi} we showed how the family of instanton based techniques can be used to estimate and reduce error floors for different decoders operating on a variety of channels. Unfortunately, instanton search is computationally prohibitive for construction of moderate length codes, and in this paper we propose another, simpler, method.

In the absence of a complete understanding of trapping sets for the SPA, the choice of forbidden subgraphs may be derived based on the understanding of trapping sets for simpler decoding algorithms as well as on intuition gained from experimental results. This is the approach we take in this paper. A basis for removing harmful trapping sets for the SPA is the observation by Chilappagari \textit{et al.} \cite{instantonPaper_shashi} that the decoding failures for various decoding algorithms and channels are closely related and that subgraphs responsible for these failures share some common underlying topological structures. These structures are either trapping sets for iterative decoding algorithms on the BSC or larger subgraphs containing these trapping sets. 

The method consists of three main steps. First, we develop a database of trapping sets for the Gallager A/B algorithm on the BSC. This database, which is called the Trapping Set Ontology (TSO)\footnote{This database of trapping sets was partially presented in \cite{ontology_vasic} and is available online at \cite{website}}, contains subgraphs that are responsible for failures of the Gallager A/B decoder and also specifies the topological relations among them. Second, based on the TSO, we determine the \textit{relative harmfulness} of different subgraphs for the SPA on the BSC by analyzing failures of the decoder on one subgraph in the presence or absence of other topologically related subgraphs. This analysis is performed repeatedly on a number of ``test'' Tanner graphs, which are intentionally constructed to either contain or be free of specific subgraphs. The relative harmfulness of a subgraph is evaluated based on its effect on the guaranteed correction capability of a code. Finally, a code is constructed so that its Tanner graph is free of the most harmful subgraphs. 

It can be seen that our construction attempts to optimize a code for the SPA on the BSC. Due to much higher complexity, similar analysis on the AWGNC is difficult. However, experimental results show that codes constructed for the BSC also perform very well on the AWGNC. It should be noted that in \cite{07DZAWN}, extensive computer simulation and hardware emulation suggest that \textit{absorbing sets} mainly contribute to error floors of codes under the SPA on the AWGNC. Since absorbing sets are combinatorially similar to trapping sets for the Gallager A/B decoder, our newly constructed codes are also free of some (and probably the most harmful) absorbing sets and hence understandably possess good error performance on the AWGNC. Although absorbing sets were invented in research that dealt with the AWGNC, their unproven harmfulness prohibit an explicit strategy to construct codes for the AWGNC. As a result, optimizing codes for the BSC in order to obtain good performance on the AWGNC remains a reasonable approach.

The rest of the paper is organized as follows. In Section \ref{sec_preliminaries}, we provide background related to LDPC codes and the necessary preliminaries for the description of the new codes. In Section \ref{sec_additive}, we propose a new class of codes based on Latin squares obtained from the additive group of a Galois field. Relations of the new codes with existing codes in the literature can be found discussed in Appendices \ref{sect_arrayCode} and \ref{sect_LinCodes}. We continue with the presentation of our Trapping Set Ontology for the Gallager A/B decoder in Section \ref{sect_ontology}. Analytical construction of a code free of trapping sets is difficult and hence we resort to an efficient search of the Tanner graph for certain subgraphs. We briefly discuss these search techniques in Section \ref{sect_search}, with more details are given in Appendix \ref{sect_search2}. In Section \ref{sect_ConsProgressive}, we describe in general the construction of a code free of certain trapping sets. We present the constructions of codes for the Gallager A/B algorithm and the SPA on the BSC in Sections \ref{sect_GAB} and \ref{sect_codesSPA}. In Section \ref{discussion}, we show the performance of several codes on the AWGNC and then conclude the paper.

\section{Preliminaries}\label{sec_preliminaries}
In this section, we introduce the definitions and notation used throughout the paper.
\subsection{LDPC Codes}
Let $\mathcal{C}$ denote an ($n,k$) LDPC code over the binary field GF(2). $\mathcal{C}$ is defined by the null space of $H$, an $m\times n$ \textit{parity check matrix} of $\mathcal{C}$. $H$ is the bi-adjacency matrix of $G$, a Tanner graph representation of $\mathcal{C}$. $G$ is a bipartite graph with two sets of nodes: $n$ variable (bit) nodes $V = \{1, 2,\ldots, n\}$ and $m$ check nodes $C = \{1, 2,\ldots ,m\}$. A vector ${\bf y}= (y_1,y_2,\ldots,y_n)$ is a codeword if and only if ${\bf y} H^\mathrm{T} = 0$, where $H^\mathrm{T}$ is the transpose of $H$. The support of $\bf y$, denoted as $\mathrm{supp}(\bf y)$, is defined as the set of all variable nodes (bits) $v \in V$ such that $y_v \neq 0$. A $d_v$-left-regular LDPC code has a Tanner graph $G$ in which all variable nodes have degree $d_v$. Similarly, a $d_c$-right-regular LDPC code has a Tanner graph $G$ in which all check nodes have degree $d_c$. A ($d_v, d_c$) regular LDPC code is $d_v$-left-regular and $d_c$-right-regular. Such a code has rate $R \geq 1 - d_v/d_c$ \cite{ldpcBook_gallager}. The degree of a variable node (check node, resp.) is also referred to as the left degree (right degree, resp.) or the column weight (row weight, resp.). The length of the shortest cycle in the Tanner graph $G$ is called the girth $g$ of $G$. 
\subsection{Permutation Matrices from Latin Squares}
A permutation matrix is a square binary matrix that has exactly one entry 1 in each row and each column and 0's elsewhere. Our codes make use of permutation matrices that do not have 1's in common positions. These sets of permutation matrices can be obtained conveniently from Latin squares.

A \textit{Latin square} of \textit{size} $q$ (or \textit{order} $q$) is a $q\times q$ array in which each cell contains a single symbol from a $q$-set $S$, such that each symbol occurs exactly once in each row and exactly once in each column. A Latin square of size $q$ is equivalent to the \textit{Cayley table} of a quasigroup $\mathcal{Q}$ on $q$ elements (see \cite[pp. 135--152]{comDesignBook} for details).

For mathematical convenience, we use elements of $\mathcal{Q}$ to index the rows and columns of Latin squares and permutation matrices. Let $\mathcal{L}={[l_{i,j}]}_{i,j\in \mathcal{Q}}$ denote a Latin square defined on the Cayley table of a quasigroup ($\mathcal{Q},\oplus$) of order $q$. We define $f$, an injective map from $\mathcal{Q}$ to $\mbox{Mat}(q,q,\mbox{GF}(2))$, where $\mbox{Mat}(q,q,\mbox{GF}(2))$ is the set of matrices of size $q\times q$ over GF(2), as follows:
\begin{eqnarray}\label{eq_Fdef}
f: \mathcal{Q}&\rightarrow& \mbox{Mat}(q,q,\mbox{GF}(2))\nonumber\\
\alpha &\mapsto& f(\alpha)={[m_{i,j}]}_{i,j\in \mathcal{Q}}\nonumber
\end{eqnarray}
such that:
\begin{eqnarray}
m_{i,j} &=& \left\{\begin{array}{cc}
1 &\mbox{~if~} l_{i,j} = \alpha\\
0 &\mbox{~if~} l_{i,j} \neq \alpha
\end{array}\right..\nonumber
\end{eqnarray}

According to this definition, a permutation matrix corresponding to the element $\alpha\in\mathcal{Q}$ is obtained by replacing the entries of $\mathcal{L}$ which are equal to $\alpha$ by 1 and all other entries of $\mathcal{L}$ by 0. It follows from the above definition that the images of elements of $\mathcal{Q}$ under $f$ give a set of $q$ permutation matrices that do not have 1's in common positions. This definition naturally associates a permutation matrix to an element $\alpha\in\mathcal{Q}$ and simplifies the derivation of parity check matrices that satisfy the RC constraint, as will be demonstrated in the next section.
\begin{ex}
Let $\mathcal{Q}$ be a quasigroup of order 4 with the following Cayley table:
\begin{eqnarray}
\begin{array}{c|cccc}
\oplus&0&1&2&3\\
\hline
0&0&1&2&3\\
1&1&0&3&2\\
2&2&3&1&0\\
3&3&2&0&1
\end{array}\nonumber
\end{eqnarray}
The Latin square obtained from the Cayley table of $\mathcal{Q}$ is:
\begin{eqnarray}
\mathcal{L} = \left[\begin{array}{cccc}
0&1&2&3\\
1&0&3&2\\
2&3&1&0\\
3&2&0&1
\end{array}\right].\nonumber
\end{eqnarray}

The injective map $f$ sends elements of $\mathcal{Q}$ to four permutation matrices:
\begin{eqnarray}
f(0) &=& \left[\begin{array}{cccc}
1&0&0&0\\
0&1&0&0\\
0&0&0&1\\
0&0&1&0\\
\end{array}\right],
f(1) = \left[\begin{array}{cccc}
0&1&0&0\\
1&0&0&0\\
0&0&1&0\\
0&0&0&1\\
\end{array}\right],\nonumber\\
f(2) &=& \left[\begin{array}{cccc}
0&0&1&0\\
0&0&0&1\\
1&0&0&0\\
0&1&0&0\end{array}\right],
f(3)= \left[\begin{array}{cccc}
0&0&0&1\\
0&0&1&0\\
0&1&0&0\\
1&0&0&0\\
\end{array}\right].\nonumber
\end{eqnarray}
\end{ex}

\subsection{LDPC Codes as Arrays of Permutation Matrices}\label{generalConstruction}
The definition of an LDPC code whose parity check matrix is an array of permutation matrices is now straightforward.
Let $\mathcal{W} = {[w_{i,j}]}_{1\leq i\leq \mu,1\leq j\leq \eta}$ be an $\mu\times \eta$ matrix over a quasigroup $\mathcal{Q}$, i.e.,
\begin{eqnarray}\label{eq_Adef}
\mathcal{W} = \left[\begin{array}{cccc}
w_{1,1}&w_{1,2}&\cdots&w_{1,\eta}\\
w_{2,1}&w_{2,2}&\cdots&w_{2,\eta}\\
\vdots&\vdots&\ddots&\vdots\\
w_{\mu,1}&w_{\mu,2}&\cdots&w_{\mu,\eta}
\end{array}
\right].
\end{eqnarray}

With some abuse of notation, let $\mathcal{H} = f(\mathcal{W}) = {[f(w_{i,j})]}$ be an array of permutation matrices, obtained by replacing elements of $\mathcal{W}$ with their images under $f$, i.e., 
\begin{eqnarray}\label{eq_Hdef}
\mathcal{H} = \left[\begin{array}{cccc}
f(w_{1,1})&f(w_{1,2})&\cdots&f(w_{1,\eta})\\
f(w_{2,1})&f(w_{2,2})&\cdots&f(w_{2,\eta})\\
\vdots&\vdots&\ddots&\vdots\\
f(w_{\mu,1})&f(w_{\mu,2})&\cdots&f(w_{\mu,\eta})\\
\end{array}
\right].
\end{eqnarray}
Then $\mathcal{H}$ is a binary matrix of size $\mu q\times \eta q$. The null space of $\mathcal{H}$ gives an LDPC code $\mathcal{C}$ of length $\eta q$. The column weight and row weight of $\mathcal{C}$ are $d_v = \mu$ and $d_c = \eta$, respectively.

We remark that different permutations of rows and columns of the Latin square $\mathcal{L}$ result in different sets of permutation matrices. These  sets of permutation matrices result in different permutations of $\mathcal{H}$ in (\ref{eq_Hdef}). Since permuting rows and columns of $\mathcal{H}$ only leads to the relabeling of the variable nodes and check nodes of the corresponding Tanner graph, different permutations of rows and columns of the Latin square $\mathcal{L}$ result in the same code. Therefore, a code is completely specified by a quasigroup $\mathcal{Q}$ along with a matrix over $\mathcal{Q}$.

\section{Structured LDPC Codes from Galois Fields of Permutation Matrices}\label{sec_additive}
The codes in this section are obtained when $\mathcal{Q}$ is the additive group of a Galois field. When $\mathcal{Q}$ is the multiplicative group of a Galois field, the codes proposed in \cite{qcFiniteField_lanLin} are obtained. We discuss this class of codes in Appendix \ref{sect_LinCodes}. The codes in this section also contain array LDPC codes \cite{arrayCode_fan} when the Galois field is a prime field, as shown in Appendix \ref{sect_arrayCode}.
\subsection{Galois Fields of Permutation Matrices}

Consider a Galois field GF($q$), where $q = p^\vartheta$, $\vartheta \in \mathbb{Z}$ and $p$ is prime. Let $\alpha$ be a primitive element of GF($q$). The powers of $\alpha$, $\alpha^{-\infty}\triangleq 0, \alpha^{0}= 1, \alpha, \alpha^{2},\ldots,\alpha^{q-2}$, give all $q$ elements of GF($q$) and $\alpha^{q-1} = 1$. Let $\mathcal{L}={[l_{i,j}]}_{i,j\in \mathcal{Q}}$ denote a Latin square defined by the Cayley table of $(\mathcal{Q},\oplus)$ where $\mathcal{Q} = \{0,1,\alpha,\ldots,\alpha^{q-2}\}$ and $\oplus$ is the subtractive operation of GF($q$), i.e., $l_{i,j} = i-j$. Although the rows and columns of $\mathcal{L}$ can be indexed arbitrarily, for simplicity we assume that they are indexed from top to bottom and left to right with increasing powers of $\alpha$. Let $\mathcal{M} = \{M_{-\infty}, M_0, M_1, \ldots, M_{q-2}\}$ be the set of images of elements of $\mathcal{Q}$ under $f$, i.e., $M_t = {[m^{(t)}_{i,j}]}_{i,j\in \mathcal{Q}} = f(\alpha^t)$. It is easy to see that $M_{-\infty}=I$, the $q\times q$ identity matrix. To show that $\mathcal{M}$ forms a field isomorphic to GF($q$) under the matrix operations defined below, we give the following propositions.
\begin{prop}\label{prop_addi}
For all $t_1, t_2 \in\mathbb{Z}$, $f(\alpha^{t_1}+\alpha^{t_2}) = M_{t_1}M_{t_2}$.
\end{prop}

\IEEEproof
Let $\Xi = [\xi_{i,j}]_{i,j\in\mathcal{Q}} = M_{t_1}M_{t_2}$ then
\begin{eqnarray}
\xi_{i,j} = \sum_{r}{m^{(t_1)}_{i,r}m^{(t_2)}_{r,j}}.\nonumber
\end{eqnarray}
Since $M_{t_1}$ and $M_{t_2}$ are permutation matrices, $\Xi$ is a permutation matrix. Assume that $\xi_{i,j} = 1$. Then there exists $r\in\mathcal{Q}$ such that $m^{(t_1)}_{i,r} = m^{(t_2)}_{r,j} = 1$. This indicates that $i-r = \alpha^{t_1}$ and $r-j = \alpha^{t_2}$. Adding, $i-j = \alpha^{t_1}+\alpha^{t_2}$ and hence $\Xi = M_{t_1}M_{t_2} = f(\alpha^{t_1}+\alpha^{t_2})$.
\endIEEEproof
\begin{cor}
${M_t}^p = I$, $\forall t$.
\end{cor}
\begin{prop}\label{pmq_prop}\label{prop_multi}
For all $t\geq0$, $M_{t+1} = PM_{t}Q$, where $P$ is a $q\times q$ permutation matrix given as
\begin{eqnarray}
P = \left[\begin{array}{cccccc}
1&0&0&\cdots&0&0\\
0&0&0&\cdots&0&1\\
0&1&0&\cdots&0&0\\
0&0&1&\cdots&0&0\\
\vdots&\vdots&\vdots&\ddots&\vdots&\vdots\\
0&0&0&\cdots&1&0\\
\end{array}
\right],
\end{eqnarray}
and $Q = P^\mathrm{T}$, the transpose of $P$.
\end{prop}
\IEEEproof
Consider two matrices $M_t = f(\alpha^t)$ and $M_{t+1} = f(\alpha^{t+1})$ for some $t \geq 0$. Assume that $m^{(t)}_{i,j} = 1$, then $l_{i,j} = i-j = \alpha^t$. Consequently, $l_{\alpha i,\alpha j} = \alpha(i - j) = \alpha^{t+1}$ and $m^{(t+1)}_{\alpha i,\alpha j} = 1$. Therefore, we can obtain $M_{t+1}$ from $M_t$ by performing the following two operations:
\begin{itemize}
\item Cyclic permutation of the last $q - 1$ rows of $M_t$, and
\item Cyclic permutation of the last $q - 1$ columns of the resulting matrix.
\end{itemize}
It is now clear that $M_{t+1} = P M_t Q$.
\endIEEEproof

Define the addition $\boxplus$ and the multiplication $\boxdot$ on $\mathcal{M}$ as:
\begin{eqnarray}
M_{t_1}\boxplus M_{t_2} &=& M_{t_1}M_{t_2},\nonumber\\
M_{t_1}\boxdot M_{t_2} &=& P^{t_2}M_{t_1}Q^{t_2}\nonumber\\
&=& P^{t_1}M_{t_2}Q^{t_1}\nonumber
\end{eqnarray}
then $\mathcal{M}$ together with $\boxplus$ and $\boxdot$ form a field isomorphic to GF($q$).

\textit{Remark:} Assume that the rows and columns of $\mathcal{L}$ are indexed arbitrarily. Let $(\alpha^{i_1},\alpha^{i_2},\ldots,\alpha^{i_q})$ be indices of the rows of $\mathcal{L}$ from top to bottom and let $(\alpha^{j_1},\alpha^{j_2},\ldots,\alpha^{j_q})$ be indices of the columns of $\mathcal{L}$ from left to right. Proposition \ref{pmq_prop} holds if $P$ and $Q$ are chosen so that the indices of the rows (from top to bottom) and the columns (from left to right) of $P\mathcal{L}Q$ are $(\alpha^{i_1+1},\alpha^{i_2+1},\ldots,\alpha^{i_q+1})$ and $(\alpha^{j_1+1},\alpha^{j_2+1},\ldots,\alpha^{j_q+1})$, respectively.
\subsection{LDPC Codes from Galois Fields of Permutation Matrices}
Define $\mathcal{W}$ and $\mathcal{H}$ as in (\ref{eq_Adef}) and (\ref{eq_Hdef}), where $(\mathcal{Q},\oplus)$ is the set $\{0,1,\alpha,\ldots,\alpha^{q-2}\}$ together with the subtractive operation of GF($q$). The following theorem gives a necessary and sufficient condition on $\mathcal{\mathcal{W}}$, such that the Tanner graph corresponding to $\mathcal{H}$ has girth at least 6.
\begin{te}[Cross-addition Constraint]\label{te_crossAddi}
The Tanner graph corresponding to $\mathcal{H}$ contains no cycle of length four iff $w_{i_1,j_1}+w_{i_2,j_2}\neq w_{i_1,j_2}+w_{i_2,j_1}$ for any $1\leq i_1,i_2 \leq \mu$; $1\leq j_1,j_2 \leq \eta$; $i_1\neq i_2$; $ j_1\neq j_2$.
\end{te}\IEEEproof
The Tanner graph corresponding to $\mathcal{H}$ contains at least one cycle of length four if and only if there exist two rows of $\mathcal{H}$ that have 1's in at least two common positions. Treat $\mathcal{H}$ as a matrix over $\mathbb{R}$ and let $\Xi = \mathcal{H}\mathcal{H}'$. Then $\Xi$ is a matrix over $\mathbb{R}$. $\mathcal{H}$ contains two rows that have 1's in at least two common positions if and only if $\Xi$ contains at least one non-diagonal component $\varpi>1$. Since $\mathcal{H}$ is an array of matrices, $\Xi$ is also an array of matrices. Also, $f(\alpha^t)$ is a permutation matrix, so its transpose is its inverse and is $f(-\alpha^t)$ (by Proposition \ref{prop_addi}). Therefore, $\Xi = {[\xi_{i_1,i_2}]}_{1\leq i_1\neq i_2\leq\mu}$ where
\begin{eqnarray}
\xi_{i_1,i_2} = \sum_{r = 1}^{\eta}{f(w_{i_1,r})f(-w_{i_2,r})}.\nonumber
\end{eqnarray}
Since $f(w_{i_1,r})f(-w_{i_2,r})\in\mathcal{M}$, $\xi_{i_1,i_2}$ contains an element $\varpi>1$ if and only if there exist $j_1\neq j_2$ such that :
\begin{eqnarray}
f(w_{i_1,j_1})f(-w_{i_2,j_1}) &=& f(w_{i_1,j_2})f(-w_{i_2,j_2})\nonumber\\
\Leftrightarrow w_{i_1,j_1} - w_{i_2,j_1} &=& w_{i_1,j_2} - w_{i_2,j_2}\nonumber\\
\Leftrightarrow w_{i_1,j_1} + w_{i_2,j_2}  &=& w_{i_1,j_2}+w_{i_2,j_1}\nonumber
\end{eqnarray}
\endIEEEproof

It can be seen that the construction of an LDPC code with girth at least 6 from a Galois field of permutation matrices reduces to finding a matrix $\mathcal{W}$ that satisfies the cross-addition constraint. 
\begin{ex} 
It can be noticed that a Latin square obtained from the Cayley table of the multiplicative group of GF($q$) satisfies the cross-addition constraint. The cross-addition constraint is still satisfied if a row and a column of all zero are appended to such a Latin square. Therefore, one form of $\mathcal{W}$ that satisfies the cross-addition constraint is given by
\begin{eqnarray}\label{eq_Wdef}
\mathcal{W} = \left[\begin{array}{ccccc}
0&0&0&\!\cdots&0\\
0&1&\alpha&\cdots&\alpha^{q-2}\\
0&\alpha&\alpha^2&\cdots&1\\
\vdots&\vdots&\vdots&\ddots&\vdots\\
0&\alpha^{q-2}&1&\cdots&\alpha^{q-3}\\
\end{array}\right].
\end{eqnarray}

Let $\mathcal{H} = f(\mathcal{W})$. From Proposition \ref{pmq_prop}, it follows that $\mathcal{H}$ has the following structure:
\begin{eqnarray}\label{eq_Hdef2}
\mathcal{H} = \left[
\begin{array}{ccccc}
I&I&I&\cdots&I\\
I&M_0&M_1&\cdots&M_{q-2}\\
I&M_1&M_2&\cdots&M_0\\
\vdots\!\!&\vdots&\vdots&\ddots&\vdots\\
I&M_{q-2}&M_0&\cdots&M_{q-3}\\
\end{array}
\right],
\end{eqnarray}
where $M_t = P^tM_0Q^t$ and $I$ is the $q\times q$ identity matrix. $\mathcal{H}$ is an array of permutation matrices from $\mathcal{M}$ and is a $q^2\times q^2$ matrix over GF($q$) with both row and column weights $q$. Since $\mathcal{W}$ satisfies the cross-addition constraint, the Tanner graph corresponding to $\mathcal{H}$ contains no cycle of length 4.

For any pair ($\gamma,\rho$) of positive integers with $1\leq \gamma,\rho \leq q$, let $H$ be a $\gamma\times\rho$ subarray of $\mathcal{H}$. Then $H$ is a $\gamma q\times \rho q$ matrix over GF(2) which is also free of cycles of length 4. $H$ has constant column weight $d_v = \gamma$ and row weight $d_c = \rho$. The null space of $H$ gives a regular structured LDPC code $\mathcal{C}$ of length $\rho q$. It can be shown that the rank of $H$ is $q\gamma - \gamma+1$, and hence $\mathcal{C}$ has rate $R = \frac{q-\gamma}{q}+\frac{\gamma-1}{q^2}$.
\end{ex}
\subsection{Remarks}
For any parity check matrix $\mathcal{H}'$ which is an array of permutation matrices, we can permute the rows and columns to obtain $\mathcal{H}$ such that the topmost and leftmost permutation matrices of $\mathcal{H}$ are identity matrices. The matrix $\mathcal{H}$ is the image of a matrix $\mathcal{W}$ under $f$, where entries on the first row and first column of $\mathcal{W}$ are $0\in\mathrm{GF}(q)$. Therefore, in the rest of the paper, we only consider matrices $\mathcal{W}$ of which elements on the first row and on the first column are zeros. For simplicity, we denote $\mathcal{U}$ as the submatrix of $\mathcal{W}$ such that
\begin{eqnarray}
\mathcal{W} = 
\begin{bmatrix}
\mathbf{0}&\mathbf{0}\\
\mathbf{0}&{\mathcal{U}}
\end{bmatrix}
\end{eqnarray}
and then write $\mathcal{H} = f(\mathcal{W}) = \bar{f}(\mathcal{U})$.

It can be seen that the notion of Latin squares provides a general and elegant description for a wide variety of structured LDPC codes whose parity check matrices are arrays of permutation matrices. For the codes described in this section, the permutation matrices are more general than circulant permutation matrices as the circulant property for our codes holds on indices understood as elements of GF($q$). Specifically, the permutation matrix corresponding to $\alpha^t$ sends the indices $(0, 1, \alpha,\ldots,\alpha^{q-2})$ to $(0+\alpha^t, 1+\alpha^t, \alpha+\alpha^t,\ldots,\alpha^{q-2}+\alpha^t)$. 

In Appendix \ref{sect_arrayCode}, we show that the class of codes described in this section includes array LDPC codes \cite{arrayCode_fan}. In particular, let $q$ be a prime then an array LDPC code is a subarray $H_\mathrm{arr}$ of the binary matrix $\mathcal{H}_\mathrm{arr}$ that is obtained by permuting rows and columns of $\mathcal{H}$ in (\ref{eq_Hdef2}) in a certain way. Note that similar to $\mathcal{H}$ in (\ref{eq_Hdef2}), $\mathcal{H}_\mathrm{arr}$ is also a $q\times q$ array of permutation matrices. In \cite{highRateG8_vasic,shortenedArrayCode_milenkovic}, a method is given to construct a shortened array LDPC code of large girth by selecting certain blocks of columns of $H_\mathrm{arr}$ to form the parity check matrix. Assume that $H^\mathrm{(s)}_\mathrm{arr}$ is such a parity check matrix then $H^\mathrm{(s)}_\mathrm{arr}$ is a subarray of $H_\mathrm{arr}$ and is also a subarray of $\mathcal{H}_\mathrm{arr}$. This approach utilizes the fact that the Tanner graph representation of $\mathcal{H}_\mathrm{arr}$ is free of four cycles and hence the Tanner graph representation of $H^\mathrm{(s)}_\mathrm{arr}$ is also free of four cycles. However, starting from on a predefined matrix $H_\mathrm{arr}$ is not a good solution in terms of code rate. This is because the fact that $H^\mathrm{(s)}_\mathrm{arr}$ is a subarray of $\mathcal{H}_\mathrm{arr}$ can also be understood as a constraint on $H^\mathrm{(s)}_\mathrm{arr}$ and therefore one might expect this constraint to reduce the code rate. The description of the codes proposed in this section along with the cross-addition constraint allow the construction of a parity check matrix in which the above-mentioned constraint is eliminated. This method of construction will be presented in Section \ref{subsect_construct}. Since the constraint is eliminated, the construction usually results in codes with higher rates than these of shortened array codes. In this paper, we use this construction method TSO (presented in the next section) to obtain codes with low error floors.  


\section{Trapping Set Ontology}\label{sect_ontology}
In this section, we describe our database of trapping sets known as the Trapping Set Ontology. This database will be used as a guideline for the construction of codes free of small trapping sets to be presented in subsequent sections. We start with a brief discussion of trapping sets and related objects.

\subsection{Trapping Sets}
A trapping set for an iterative decoding algorithm is defined as a non-empty set of variable nodes in a Tanner graph $G$ that are not eventually corrected by the decoder \cite{errorFloor_richarson}. A set of variable nodes $\bf T$ is called an ($a,b$) trapping set if it contains $a$ variable nodes and the subgraph induced by these variable nodes has $b$ odd degree check nodes.

For transmission over the BEC, trapping sets are characterized combinatorially and are known as stopping sets \cite{bec_di}. For transmission over the AWGNC, no explicit combinatorial characterization of trapping sets has been found. In the case of the BSC, when decoding with the Gallager A/B algorithm, or the bit flipping (serial or parallel) algorithms, then trapping sets are partially characterized under the notion of fixed sets. By partially, we mean that these combinatorial objects form a subclass of trapping sets, but not all trapping sets are fixed sets. Fixed sets have been studied extensively in a series of conference papers \cite{errorFloorBSC_shashi, clw3paper, guaranteedError_shashic,cover_milos}. They have been proven to be the cause of error floor in the decoding of LDPC codes under the Gallager A/B algorithm and the bit flipping algorithms. For the sake of completeness, we give the definition of a fixed set as well as the necessary and sufficient conditions for a set of variable nodes to form a fixed set.

Consider an iterative decoder on the BSC. Assume the transmission of an all-zero codeword\footnote{The all-zero-codeword assumption can be applied if the channel is output symmetric and the decoding algorithms satisfied certain symmetry conditions (see Definition 1 and Lemma 1 in \cite{capacityLDPCmp_richarson}). The Gallager A/B algorithm, the bit flipping algorithms and the SPA all satisfy these symmetry conditions.
}. With this assumption, a variable node is correct if it is 0 and corrupt if it is 1. Let ${\bf y} = (y_1, y_2,\ldots,y_n)$ be the input to the decoder and let ${\bf x}^l = (x^l_1, x^l_2,\ldots, x^l_n)$ be the output vector at the $l^\mathrm{th}$ iteration. Let $\bf F(y)$ denote the set of variable nodes that are not eventually correct.
\begin{defn}
For transmission over the BSC, $\bf y$ is a fixed point of the decoding algorithm if $\mathrm{supp}({\bf y}) = \mathrm{supp}({\bf x}^l)$ for all $l$. If ${\bf F(y)} \neq \emptyset$ and $\bf y$ is a fixed point, then ${\bf F(y)} = \mathrm{supp}({\bf y})$ is a fixed set. A fixed set (trapping set) is  \textit{an elementary fixed set (trapping set)} if all check nodes in its induced subgraph have degree one or two\footnote{This classification was given in \cite{spectraTS_milenkovic}.}. Otherwise, it is a non-elementary fixed set (trapping set). 
\end{defn}

\begin{te}[\cite{guaranteedError_shashic}]\label{te_fSet} Let $\mathcal{C}$ be an LDPC code with $d_v$-left-regular Tanner graph $G$. Let $\bf T$ be a set consisting of variable nodes with induced subgraph $\bf I$. Let the check nodes in $\bf I$ be partitioned into two disjoint subsets; $\bf O$ consisting of check nodes with odd degree and $\bf E$ consisting of check nodes with even degree. Then $\bf T$ is a fixed set for the bit flipping algorithms (serial or parallel) iff : (a) Every variable node in $\bf I$ has at least $\lceil \frac{d_v}{2} \rceil$ neighbors in $\bf E$ and (b) No $\lfloor \frac{dv}{2} \rfloor+1$ check nodes of $\bf O$ share a neighbor outside $\bf I$.
\end{te}

Note that Theorem \ref{te_fSet} only states the conditions for the bit flipping algorithms. However, it is not difficult to show that these conditions also apply for the Gallager A/B algorithm.

Although it has been rigorously proven only that fixed sets are trapping sets for the Gallager A/B algorithm and the bit flipping algorithms on the BSC, it has been widely recognized in the literature that the subgraphs of these combinatorial objects greatly contribute to the error floor for various iterative decoding algorithms and channels. The instanton analysis performed by Chilappagari \textit{et al.} in \cite{instantonPaper_shashi} suggests that the decoding failures for various decoding algorithms and channels are closely related and subgraphs responsible for these failures share some common underlying topological structures. These structures are either trapping sets for iterative decoding algorithms on the BSC, of which fixed sets form a subset, or larger subgraphs containing these trapping sets. Dolecek \textit{et al.} in \cite{absorbingSet_dolecek} defined the notion of absorbing sets, which is very similar to the notion of fixed sets. By hardware emulation, they found that absorbing sets are the main cause of error floors for the SPA on the AWGNC. Various trapping sets identified by simulation (for example those in \cite{toward_ryan,nearCW_mackay}) are also fixed sets. 

From these observations, it is expected that an LDPC code will have low error floor performance if the corresponding Tanner graph does not contain subgraphs induced by fixed sets. However, it is impossible to construct an LDPC code whose Tanner graph is free of all fixed sets when the length of the code is finite. It is also well-known that imposing constraints on a Tanner graph reduces the rate of a code. Clearly, only subgraphs of some fixed sets can be avoided in the code construction. These need to be chosen carefully in order to obtain the best possible error floor performance while maximizing the code rate.

Before one can attempt to determine the fixed sets to forbid in the Tanner graph of a code,  there are two important issues that need to be addressed. First, a complete list of non-isomorphic fixed sets (up to a proper size) for a given set of code parameters (e.g., column weight and row weight) is needed. This is because the notion of an $(a,b)$ fixed set (trapping set) is not sufficient. Given a pair of positive integers $(a,b)$, there are possibly many fixed sets which induce non-isomorphic subgraphs containing $a$ variable nodes and $b$ odd degree check nodes. Second, the topological relations among subgraphs induced by fixed sets needs to be explored. The importance of these relations is threefold. First, the subgraph induced by a fixed set may be contained in the subgraph induced by another fixed set. In such case the absence of one subgraph yields to the absence of the other. Second, these relations help reduce the complexity of the search for subgraphs in a Tanner graph. Finally, these relations reduce the complexity of the analysis to determine the harmfulness of subgraphs.

In the next subsection, we present our database of fixed sets for regular column-weight-three LDPC codes with emphasis on the topological relations among them. For the sake of simplicity, we drop the term fixed sets and refer to these objects by the general term trapping sets. 

%
\subsection{Trapping Set Ontology of Column-Weight-Three Codes for the Gallager A/B Algorithm on the BSC}
\subsubsection{Graphical representation}
The induced subgraph of a trapping set (or any set of variable nodes) is a bipartite graph. In the Tanner graph (bipartite graph) representation of a trapping set, we use $\CIRCLE$ to represent variable nodes, $\blacksquare$ to represent odd degree check nodes and $\Box$ to represent even degree check nodes. There exists an alternate graphical representation of trapping sets which allows their topological relations to be established more conveniently. This graphical representation is based on the incidence structure of lines and points. In combinatorial mathematics, an incidence structure is a triple $\mathscr{(P,L,I)}$ where $\mathscr{P}$ is a set of ``points'', $\mathscr{L}$ is a set of ``lines'' and $\mathscr{I}\subseteq \mathscr{P}\times \mathscr{L}$ is the incidence relation. The elements of $\mathscr{I}$ are called flags. If $\mathpzc{(p,l)}\in \mathscr{I}$, we say that point $\mathpzc{p}$ ``lies on'' line $\mathpzc{l}$. In this lines and points (henceforth line-point) representation of trapping sets, variable nodes correspond to lines and check nodes correspond to points. A point is shaded black if it has an odd number of lines passing through it, otherwise it is shaded white. An $(a, b)$ trapping set is thus an incidence structure with $a$ lines and $b$ black shaded points. To differentiate among $(a,b)$ trapping sets that have non-isomorphic induced subgraphs when necessary, we index $(a,b)$ trapping sets in an arbitrary order and assign the notation $(a,b)\{i\}$ to the $(a,b)$ trapping set with index $i$. 

Depending on the context, a trapping set can be understood as a set of variable nodes in a given code with a specified induced subgraph or it can be understood as a specific subgraph independent of a code. To differentiate between these two cases, we use the letter $\bf T$ to denote a set of variable nodes in a code and use the letter $\mathcal{T}$ to denote a type of trapping set which corresponds to a specific subgraph. If the induced subgraph of a set of variable nodes $\bf T$ in the Tanner graph of a code $\mathcal{C}$ is isomorphic to the subgraph of $\mathcal{T}$ then we say that $\bf T$ is a $\mathcal{T}$ trapping set or that $\bf T$ is a trapping set of type $\mathcal{T}$. $\mathcal{C}$ is said to contain $\mathcal{T}$ trapping set(s).

\begin{ex}
The $(5,3)\{1\}$ trapping set $\mathcal{T}_1$ is a union of a six cycle and an eight cycle, sharing two variable nodes. The Tanner graph representation of $\mathcal{T}_1$ is shown in  Fig. \ref{fig_grapRep}\subref{531_t}. The set of odd degree check nodes is $\{c_7,c_8,c_9\}$. These check nodes are represented by black shaded squares. In the line-point representation of $\mathcal{T}_1$ which is shown in  Fig. \ref{fig_grapRep}\subref{531_lp}, $c_7,c_8$ and $c_9$ are represented by black shaded points. These points are the only points that lie on a single line. The five variable nodes $v_1,v_2,\ldots, v_5$ are represented by black shaded circles in  Fig. \ref{fig_grapRep}\subref{531_t}. They correspond to the five lines in  Fig. \ref{fig_grapRep}\subref{531_lp}. As an example, the column-weight-three MacKay random code of length 4095 \cite{macKayCode} has 19617 sets of variable nodes whose induced subgraphs are isomorphic to the subgraph of $\mathcal{T}_1$. These sets of variable nodes are $(5,3)\{1\}$ trapping sets.

\begin{figure}
\centering
\subfigure[] 
{
    \label{531_t}

\includegraphics[height = 0.9 in]{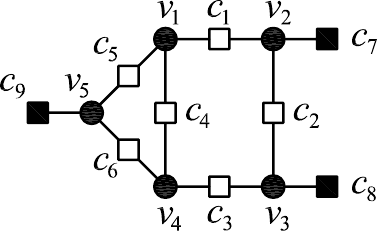}
}
\hspace{0.2in}
\subfigure[] 
{
    \label{531_lp}

\includegraphics[height = 0.9 in]{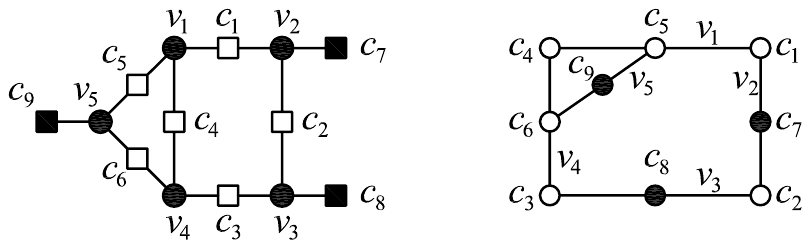}
}
\caption{Graphical representation of the $(5,3)\{1\}$ trapping set: \subref{531_t} Tanner graph representation, \subref{531_lp} Line-point representation.}
\label{fig_grapRep}
\end{figure}
\end{ex}

\textit{Remark:} To avoid confusion between the graphical representations of trapping sets, we note that the Tanner graph representation of a trapping set always contains $\Box$ or $\blacksquare$. The line-point representation never contains $\Box$ or $\blacksquare$. In the remainder of this paper, we only use the line-point representation.

\subsubsection{Topological relation}
The following definition gives the topological relations among trapping sets.
\begin{defn}
A trapping set $\mathcal{T}_2$ is a successor of a trapping set $\mathcal{T}_1$ if there exists a proper subset of variable nodes of $\mathcal{T}_2$ that induce a subgraph isomorphic to the induced subgraph of $\mathcal{T}_1$. If $\mathcal{T}_2$ is a successor of $\mathcal{T}_1$ then $\mathcal{T}_1$ is a parent of $\mathcal{T}_2$. Furthermore, $\mathcal{T}_2$ is a direct successor of $\mathcal{T}_1$ if it does not have a parent $\mathcal{T}_3$ which is a successor of $\mathcal{T}_1$. 
\end{defn}

The topological relation between $\mathcal{T}_1$ and $\mathcal{T}_2$ is solely dictated by the topological properties of their subgraphs. In the Tanner graph of a code $\mathcal{C}$, the presence of a trapping set $\bf T_1$ does not indicate the presence of a trapping set $\bf T_2$. If $\bf T_1$ is indeed a subset of a trapping set $\bf T_2$ in the Tanner graph of $\mathcal{C}$ then we say that $\bf T_1$ \textit{generates} $\bf T_2$, otherwise we say that $\bf T_1$ does not generate $\bf T_2$.

\subsubsection{Family tree of trapping sets}
Theorem \ref{te_fSet} implies that every trapping set $\mathcal{T}$ contains at least a cycle. To show this, assume that $\mathcal{T}$ is a trapping set that does not contain a cycle then the induced subgraph of $\mathcal{T}$ is a tree. Take any variable node as the root of the tree then the variable nodes which are neighboring to the leaf nodes with largest depth have only one check node with degree greater than 1. Therefore these variable nodes have no less odd degree check nodes than even degree check nodes. This indicates that $\mathcal{T}$ is not a trapping set, which is a contradiction. Consequently, all trapping sets can be obtained by adjoining variable nodes to cycles. Note that any cycle is a trapping set for regular column-weight-three codes.
\begin{figure}[h]
\centering
\subfigure[] 
{
    \label{merge531}

\includegraphics[height = 0.8 in]{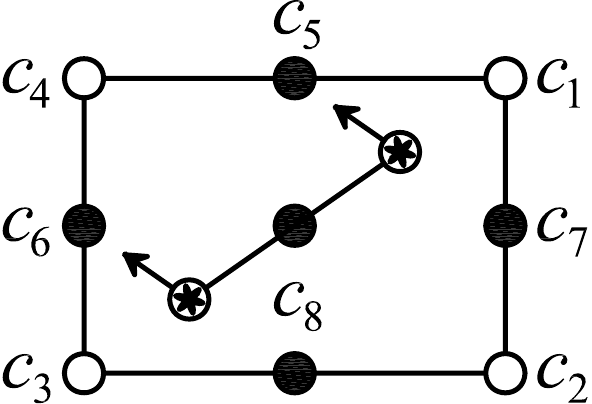}
}
\hspace{0.3in}
\subfigure[] 
{
    \label{merge532}

\includegraphics[height = 0.8 in]{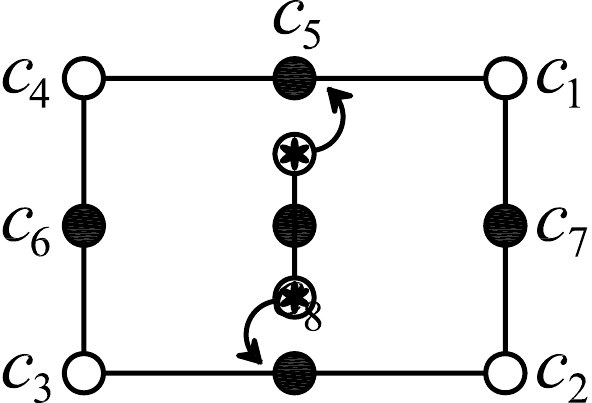}
}
\caption{Obtaining $(5,3)$ trapping sets by adding a line to the $(4,4)$ trapping set: \subref{merge531} the $(5,3)\{1\}$ trapping set and \subref{merge532} the $(5,3)\{2\}$ trapping set.}
\label{fig_merge53}
\end{figure}

We now explain how larger trapping sets can be obtained by adjoining variable nodes to smaller trapping sets. We begin with the simplest example: the evolution of ($5,3$) trapping sets from the $(4,4)$ trapping set for regular-column-weight three codes. We know that compared to the $(4,4)$ trapping set, which is an eight cycle, a ($5,3$) trapping set has one additional variable node. Therefore, if a ($5,3$) trapping set is a successor of the $(4,4)$ trapping set, then its line-point representation can be obtained by adding one additional line to the line-point representation of the $(4,4)$ trapping set. Since it is required that the addition of the new line preserves the variable node degree (or number of points lying on a line), there must be exactly three points lying on the new line. Therefore, we can consider the process of adding a new line as the merging of at least one point on the new line with certain points in the line-point representation of the $(4,4)$ trapping set. We use $\circledast$ to denote points on the line that are to be merged with points in the line-point representation of the parent trapping set. The merging is demonstrated in Fig. \ref{fig_merge53} and is explained as follows. If a black shaded point is merged with a $\circledast$ point then they become a single white shaded point. Similarly, if a white shaded point is merged with a $\circledast$ point then the result is a single black shaded point. Recall that there must be exactly three black shaded points in the line-point representation of a trapping set. In addition, every line must pass through at least two white shaded points. The only way to satisfy these two conditions is to merge two points of the new line with two black shaded points of the $(4,4)$ trapping set. There are two distinct ways to select two black shaded points, resulting in two different $(5,3)$ trapping sets.

\begin{figure}
\centering
\subfigure[] 
{
    \label{sl}

\includegraphics[height = 0.5 in, angle = 0]{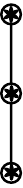}
}
\hspace{0.1in}
\subfigure[] 
{
    \label{dl}

\includegraphics[height = 0.5 in, angle = 0]{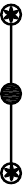}
}
\hspace{0.1in}
\subfigure[] 
{
    \label{tl}
\includegraphics[height = 0.5 in, angle = 0]{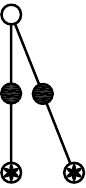}
}
\hspace{0.1in}
\subfigure[] 
{
    \label{ml}
\includegraphics[height = 0.5 in, angle = 0]{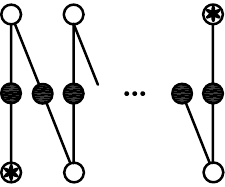}
}
\hspace{0.1in}
\subfigure[] 
{
    \label{ta1}
\includegraphics[height = 0.5 in, angle = 0]{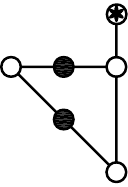}
}
\hspace{0.1in}
\subfigure[] 
{
    \label{ta2}
\includegraphics[height = 0.5 in, angle = 0]{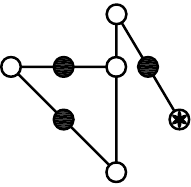}
}
\hspace{0in}
\subfigure[] 
{
    \label{ta3}
\includegraphics[height = 0.5 in, angle = 0]{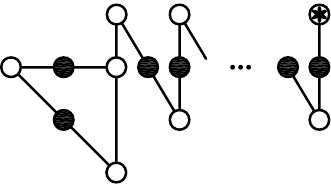}
}
\hspace{0in}
\subfigure[] 
{
    \label{sq1}
\includegraphics[height = 0.5 in, angle = 0]{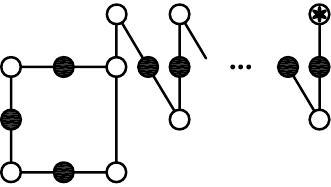}
}
\hspace{0in}
\subfigure[] 
{
    \label{sq2}
\includegraphics[height = 0.5 in, angle = 0]{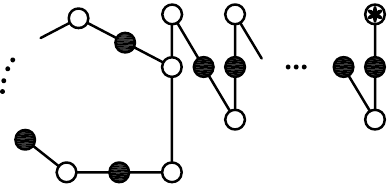}
}
\caption{Possible incidence structures formed by $u$ new lines for elementary trapping sets.}
\label{fig_adjGraph}
\end{figure}

The evolution of a trapping set for regular-column-weight three codes from its parent can now be described in a more general setting. Since every trapping set of interest is a direct successor of some trapping sets, it is sufficient to only consider the evolution of direct successors. Consider an ($a,b$) trapping set $\mathcal{T}_1$. Since $\mathcal{T}_1$ has $a$ variable nodes, its line-point representation contains $a$ lines. Each line has 3 points lying on it, with at most one point shaded black. There are $b$ black shaded points, each has an odd number of lines passing through it. The line-point representation of an ($a+u,b+z$) trapping set $\mathcal{T}_2$ can be obtained by adding $u$ lines to the line-point representation of $\mathcal{T}_1$. These $u$ new lines (and the points on them) form an incidence structure and since $\mathcal{T}_2$ is a direct successor of $\mathcal{T}_1$, this incidence structure is connected\footnote{Each incidence structure corresponds to a bipartite graph. An incidence structure is connected if the corresponding bipartite graph is connected}. For simplicity, let us only consider elementary trapping sets. Then it can be shown that the incidence structure formed by the new $u$ lines can only be one of those listed in  Fig. \ref{fig_adjGraph}. A successor trapping set $\mathcal{T}_2$ is obtained by pairwisely merging the $\circledast$ points with certain points of $\mathcal{T}_1$. We remark that non-elementary successors can be obtained in a very similar process with small additional complexity.

\subsubsection{Example} 
Let us consider regular column-weight-three LDPC codes. For simplicity, we only consider codes of girth $g=8$ and elementary trapping sets, although this example can be generalized to include codes of other girths and non-elementary trapping sets.
\begin{figure}
\centering
\subfigure[] 
{
    \label{merge641f}

\includegraphics[height = 0.5 in]{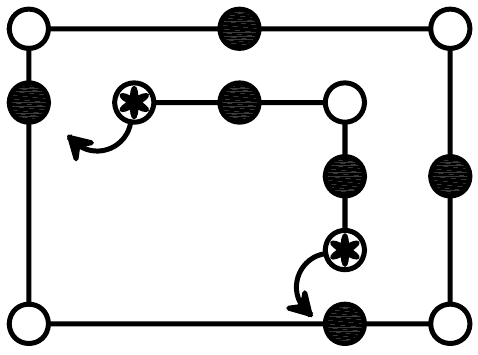}
}
\hspace{0.05in}
\subfigure[] 
{
    \label{merge642f}

\includegraphics[height = 0.5 in]{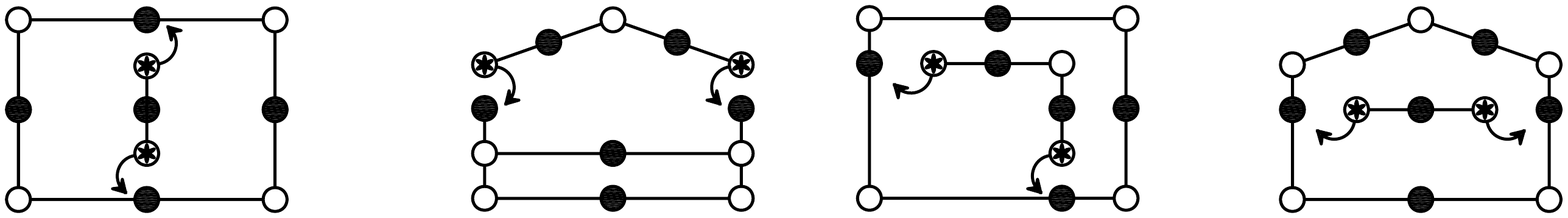}
}
\hspace{0.05in}
\subfigure[] 
{
    \label{merge642v}

\includegraphics[height = 0.5 in]{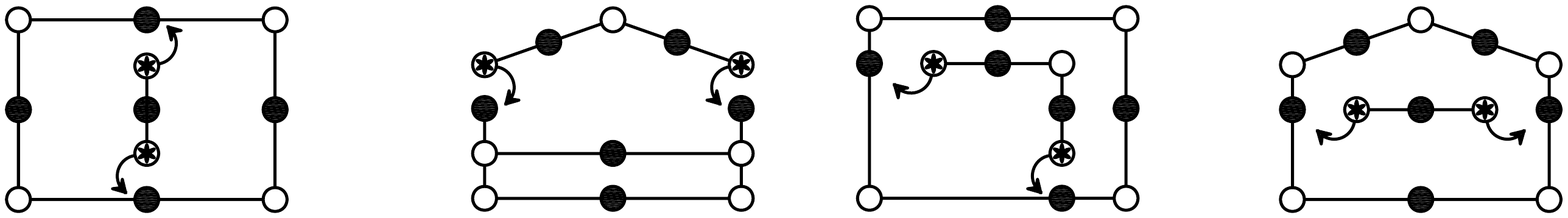}
}
\caption{Obtaining larger a trapping set by adding lines to a smaller one.}
\label{fig_merge}
\end{figure}

\begin{itemize} 

\item With the evolution of the $(5,3)\{2\}$ trapping set presented above, we show the family tree of $(a,b)$ trapping sets originating from the $(5,3)\{2\}$ trapping set with $a\leq 8$ and $b>0$ in  Fig. \ref{fig_53andChild}.
\begin{figure}
\centering
\hspace{0.0in}
\subfigure[$(5,3)\{2\}$] 
{
    \label{532_lp}

\includegraphics[height = 0.5 in]{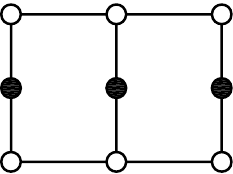}
}
\hspace{0in}
\subfigure[$(6,2)\{1\}$] 
{
    \label{621_lp}
\includegraphics[height = 0.5 in]{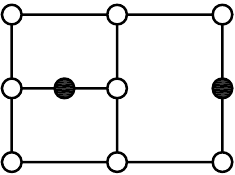}
}
\hspace{0in}
\subfigure[$(7,1)\{1\}$] 
{
    \label{711_lp}
\includegraphics[height = 0.5 in]{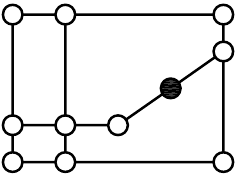}
}
\hspace{0in}
\subfigure[$(8,2)\{1\}$] 
{
    \label{821_lp}
\includegraphics[height = 0.5 in]{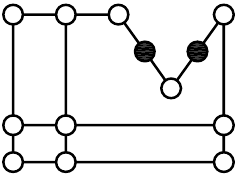}
}

\hspace{0in}
\subfigure[$(7,3)\{1\}$] 
{
    \label{731_lp}
\includegraphics[height = 0.5 in]{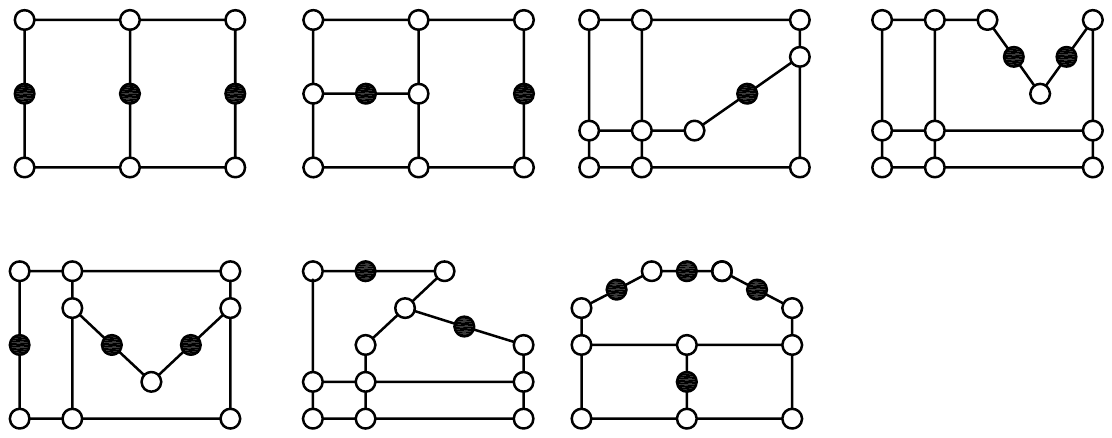}
}
\hspace{0.2in}
\subfigure[$(8,2)\{2\}$] 
{
    \label{822_lp}
\includegraphics[height = 0.5 in]{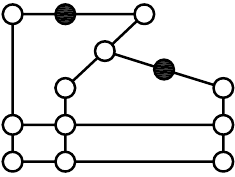}
}
\hspace{0.2in}
\subfigure[$(8,4)\{1\}$] 
{
    \label{841_lp}
\includegraphics[height = 0.5 in]{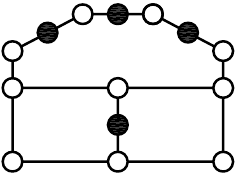}
}
\caption{The $(5,3)\{1\}$ trapping set and its successors of size less than 8 in girth 8 LDPC codes.}
\label{fig_53andChild}
\end{figure}
\begin{figure}
\centering
\hspace{0.0in}
\subfigure[$(4,4)$] 
{
    \label{441_lp}

\includegraphics[height = 0.5 in]{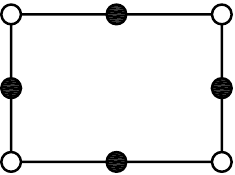}
}
\hspace{0in}
\subfigure[$(6,4)\{2\}$] 
{
    \label{642_lp}
\includegraphics[height = 0.5 in]{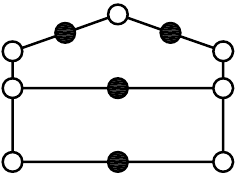}
}
\hspace{0in}
\subfigure[$(7,5)\{1\}$] 
{
    \label{751_lp}
\includegraphics[height = 0.5 in]{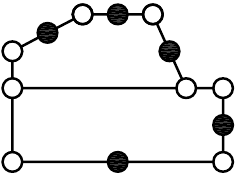}
}
\hspace{0in}
\subfigure[$(7,5)\{2\}$] 
{
    \label{752_lp}
\includegraphics[height = 0.5 in]{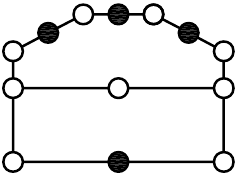}
}

\hspace{0in}
\subfigure[$(8,6)\{1\}$] 
{
    \label{861_lp}
\includegraphics[height = 0.5 in]{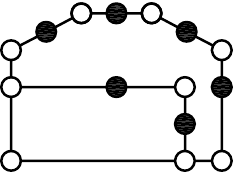}
}
\hspace{0.2in}
\subfigure[$(8,6)\{2\}$] 
{
    \label{862_lp}
\includegraphics[height = 0.5 in]{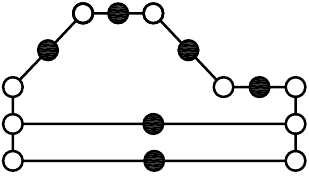}
}
\hspace{0.2in}
\subfigure[$(8,6)\{3\}$] 
{
    \label{863_lp}
\includegraphics[height = 0.5 in]{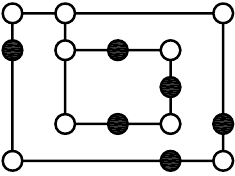}
}
\caption{The $(4,4)$ trapping set and its direct successors of size less than 8 in girth 8 LDPC codes (excluding the $(5,3)\{1\}$ and $(6,4)\{1\}$ trapping sets shown in Fig. \ref{fig_grapRep}\subref{531_lp} and \ref{fig_64andChild}\subref{641_lp}, respectively).}
\label{fig_44andChild}
\end{figure}

\item By selecting two black shaded nodes in  Fig. \ref{fig_44andChild}\subref{441_lp} and merging them with two $\circledast$ nodes in  Fig. \ref{fig_adjGraph}\subref{tl}, a $(6,4)$ trapping set can be obtained. Two distinct ways to select black shaded nodes result in two different trapping sets: the $(6,4)\{1\}$ trapping set shown in  Fig. \ref{fig_64andChild}\subref{641_lp} and the $(6,4)\{2\}$ trapping set shown in  Fig. \ref{fig_44andChild}\subref{642_lp}. The merging is demonstrated in Fig. \ref{fig_merge}\subref{merge641f} and \subref{merge642f}. The family tree of $(a,b)$ trapping sets originating from the $(6,4)\{1\}$ trapping set with $a\leq 8$ and $b>0$ is illustrated in  Fig. \ref{fig_64andChild}.
\begin{figure}
\centering
\hspace{0.02in}
\subfigure[$(6,4)\{1\}$] 
{
    \label{641_lp}

\includegraphics[height = 0.5 in]{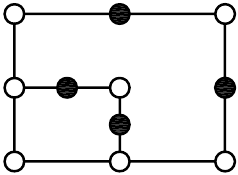}
}
\hspace{0.2in}
\subfigure[$(7,3)\{2\}$] 
{
    \label{732_lp}
\includegraphics[height = 0.5 in]{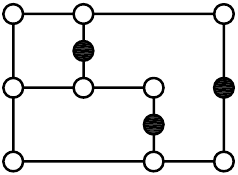}
}
\hspace{0.2in}
\subfigure[$(8,2)\{3\}$] 
{
    \label{823_lp}
\includegraphics[height = 0.5 in]{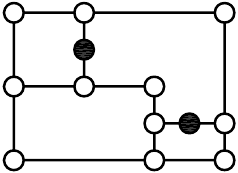}
}

\hspace{0in}
\subfigure[$(7,3)\{3\}$] 
{
    \label{733_lp}
\includegraphics[height = 0.5 in]{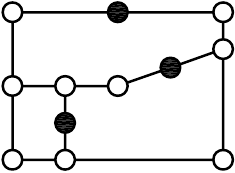}
}
\hspace{0.2in}
\subfigure[$(8,2)\{4\}$] 
{
    \label{824_lp}
\includegraphics[height = 0.5 in]{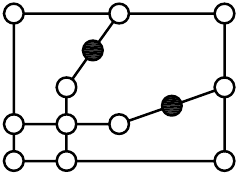}
}
\hspace{0.2in}
\subfigure[$(8,2)\{5\}$] 
{
    \label{825_lp}
\includegraphics[height = 0.5 in]{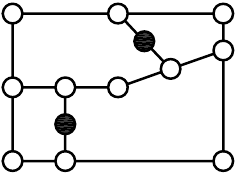}
}

\hspace{0in}
\subfigure[$(8,4)\{2\}$] 
{
    \label{842_lp}
\includegraphics[height = 0.5 in]{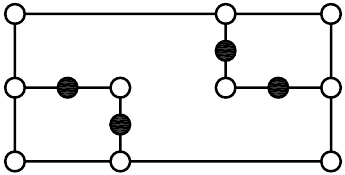}
}
\hspace{0in}
\subfigure[$(8,4)\{3\}$] 
{
    \label{843_lp}
\includegraphics[height = 0.5 in]{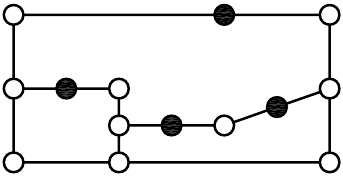}
}
\hspace{0in}
\subfigure[$(8,4)\{4\}$] 
{
    \label{844_lp}
\includegraphics[height = 0.5 in]{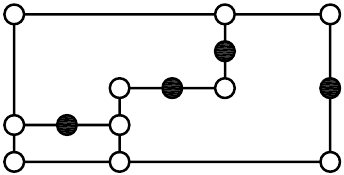}
}
\caption{The $(6,4)\{1\}$ trapping set and its successors of size less than 8 in girth 8 LDPC codes.}
\label{fig_64andChild}
\end{figure}

\item In the same manner, other direct successors of the ($4,4$) trapping set can be generated. Those $(a,b)$ trapping sets with $a\leq 8$ and $b>0$ are shown in  Fig. \ref{fig_44andChild}. For a more complete list of trapping sets from the TSO, the interested readers are referred to \cite{website}.
\end{itemize}

\textit{Remarks:} A trapping set may originate from different parents. For example, the $(6,4)\{2\}$ trapping set is not only a direct successor of the ($4,4$) trapping set but also a direct successor of the ($5,5$) trapping set. The evolution of the $(6,4)\{2\}$ trapping set from the $(5,5)$ trapping set is demonstrated in Fig. \ref{fig_merge}\subref{merge642v}.
\subsubsection{Codewords}
\begin{figure}
\centering
\subfigure[$(6,0)\{1\}$] 
{
    \label{weight6_2}

\includegraphics[height = 0.8 in, angle = 90]{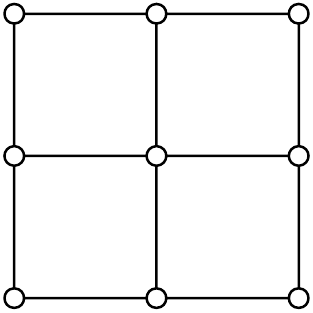}
}
\hspace{0.2in}
\subfigure[$(8,0)\{1\}$] 
{
    \label{w8_1}

\includegraphics[height = 0.8 in, angle = 90]{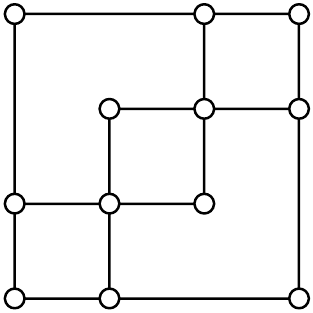}
}
\hspace{0.2in}
\subfigure[$(8,0)\{2\}$] 
{
    \label{w8_2}
\includegraphics[height = 0.8 in, angle = 90]{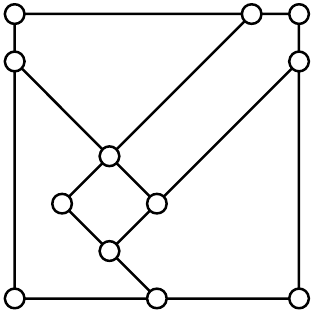}
}
\hspace{0.2in}
\subfigure[$(10,0)\{1\}$] 
{
    \label{weight10_1}
\includegraphics[height = 0.8 in, angle = 0]{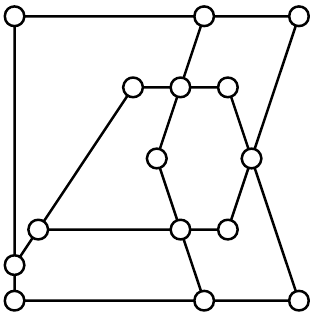}
}
\hspace{0.2in}
\subfigure[$(10,0)\{2\}$] 
{
    \label{weight10_2}
\includegraphics[height = 0.8 in, angle = 0]{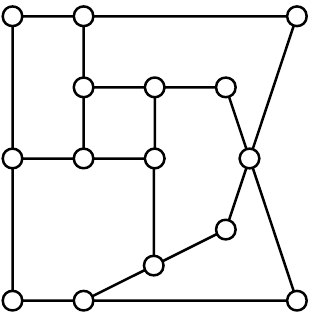}
}
\hspace{0.2in}
\subfigure[$(10,0)\{3\}$] 
{
    \label{weight10_3}
\includegraphics[height = 0.8 in, angle = 0]{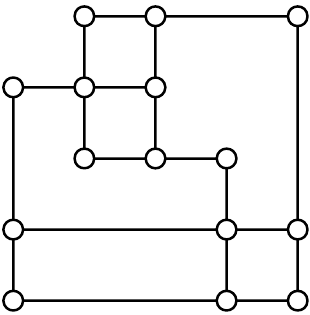}
}
\hspace{0.2in}
\subfigure[$(10,0)\{4\}$] 
{
    \label{weight10_4}
\includegraphics[height = 0.8 in, angle = 0]{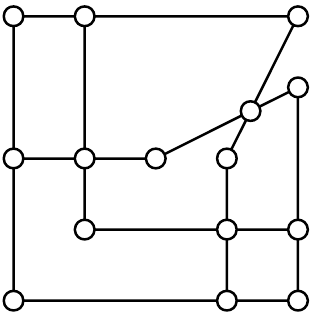}
}
\hspace{0.2in}
\subfigure[$(10,0)\{5\}$] 
{
    \label{weight10_5}
\includegraphics[height = 0.8 in, angle = 0]{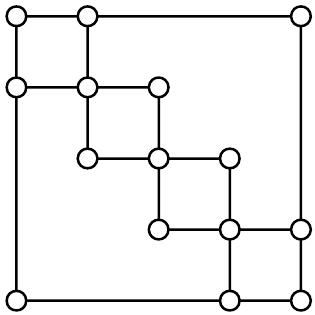}
}
\hspace{0.2in}
\subfigure[$(10,0)\{6\}$] 
{
    \label{weight10_6}
\includegraphics[height = 0.8 in, angle = 0]{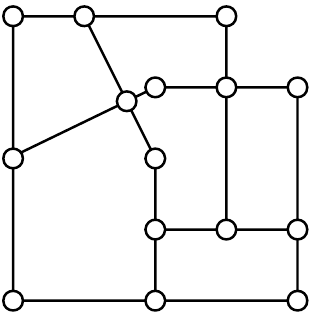}
}
\caption{All $(a,0)$ trapping sets where $a\leq 10$ in girth 8 LDPC codes.}
\label{fig_cwg8}
\end{figure}

Let $\bf y$ be a codeword of $\mathcal{C}$ and let ${\bf T} = \mathrm{supp}({\bf y})$. It is clear that $\bf T$ is an $(a,0)$ trapping set where $a = |\mathrm{supp}({\bf y})|$. Conversely, $\mathcal{C}$ contains codewords of Hamming weight $a$ if the Tanner graph of $\mathcal{C}$ contains $(a,0)$ trapping sets. It is also clear that $\mathcal{C}$ has $d_{\mathrm{min}}$ as its minimum distance if and only if (i) the Tanner graph of $\mathcal{C}$ contains no $(a,0)$ trapping set where $a<d_{\mathrm{min}}$ and (ii) the Tanner graph of $\mathcal{C}$ contains at least one $(d_{\mathrm{min}},0)$ trapping set. For regular column-weight-three codes, an $(a,0)$ trapping set is a direct successor of an $(a-1,3)$ trapping set. Consequently, the line-point representation of an $(a,0)$ trapping set is obtained by pairwisely merging three black shaded nodes in the line-point representation of an $(a-1,3)$ trapping set with three $\circledast$ nodes in  Fig. \ref{fig_adjGraph}\subref{sl}. The line-point representations of all possible $(a,0)$ trapping sets where $a\leq 10$ of girth 8 codes are shown in  Fig. \ref{fig_cwg8}.

\section{Searching for Subgraphs in a Tanner Graph}\label{sect_search}
In this section, we briefly describe the main idea behind our techniques of searching for elementary trapping sets from the TSO in the Tanner graph of a regular column-weight-three LDPC code. An efficient search of the Tanner graph for trapping sets relies on the topological relations among trapping sets defined in the TSO and/or carefully analyzing their induced subgraphs. Trapping sets are searched for in a way similar to how they have evolved in the TSO. A bigger trapping set can be found in a Tanner graph by expanding a smaller trapping set. More precisely, given a trapping set $\bf T_1$ of type $\mathcal{T}_1$ in the Tanner graph of a code $\mathcal{C}$, our techniques search for a set of variable nodes such that the union of this set with $\bf T_1$ form a trapping set $\bf T_2$ of type $\mathcal{T}_2$, where $\mathcal{T}_2$ is a successor of $\mathcal{T}_1$. Our techniques are sufficient to efficiently search for a large number of trapping sets in the TSO, especially for those to be avoided in the code constructions that we will present in subsequent sections. They can be easily expanded to search for other trapping sets as well. Details on the implementation of these techniques are given in Appendix \ref{sect_search2}. 

It is necessary to mention existing methods of searching for trapping sets in the Tanner graph of a code. It is well-known that this problem is NP hard \cite{npHard_krishnan,npHard_olgica}. Previous work on this problem includes exhaustive \cite{09WKP,exhautiveSearch_Kyung} and non-exhaustive approaches \cite{app_hiro,enum_shadi}. The main drawback of existing exhaustive approaches is high complexity. Consequently, constraints must be imposed on trapping sets and on the Tanner graph in which trapping sets are searched for. For example, the method in \cite{09WKP} can only search for $(a,b)$ trapping sets with $b\leq 2$ in a Tanner graph with less than 500 variable nodes. The complexity is much lower for non-exhaustive approaches. However, these approaches can not guarantee that all trapping sets are enumerated, and hence are not suitable for the purpose of this paper.

\section{Construction of Codes Free of Small Trapping Sets}\label{sect_ConsProgressive}
Let us begin this section by summarizing the paper up until this point. We have given the description for a general class of LDPC codes whose parity check matrices are arrays of permutation matrices obtained from Latin squares. We have also presented our database of trapping sets of regular column-weight-three codes for the Gallager A/B algorithm on the BSC. Subgraphs of these trapping sets are identified by many researchers as the main cause of error floor for various iterative decoding algorithms and channels. Methods of searching for these subgraphs in the Tanner graph of a code have also been presented. We therefore have all theoretical tools necessary to proceed to code construction.

In this section, we give a general method to construct regular LDPC codes free of a given collection of trapping sets. More precisely, codes are constructed so that their Tanner graphs are free of a given collection of subgraphs from the TSO. Therefore, in this context, an $(a,b)\{i\}$ trapping set should be understood as a specific subgraph and not as a set of non-eventually correct variable nodes. It is important to note that our method of constructing codes free of small trapping sets can be applied to any class of codes, and not just the new class of codes proposed in this paper. For example, the progressive edge growth (PEG) method \cite{peg_hu} can be modified to construct a random code whose Tanner graph is free of certain subgraphs. Similarly, the method of progressively constructing a Tanner graph described below can be applied to construct any code whose parity check matrix is an array of permutation matrices. However, we restrict ourselves to construct codes defined in Section \ref{sec_additive} in this paper, for the sole purpose of demonstrating the excellent behavior of this newly proposed class of codes.

We organize our discussion by considering two separate problems: determining a collection of forbidden subgraphs, i.e., which subgraphs that should be avoided in the Tanner graph and (ii) constructing a Tanner graph which is free of a given collection of subgraphs. Let us begin with the second problem.

\subsection{Construction of a Code by Progressively Building the Tanner Graph}\label{subsect_construct}
We give a progressive construction of a $(d_v,d_c)$ regular LDPC code whose parity check matrix is an array of permutation matrices. Our construction algorithm is inspired by the PEG algorithm \cite{peg_hu} and the method in \cite{highRateG8_vasic}. Let $\mathcal{C}$ be a $(\gamma,\rho)$ regular LDPC code whose parity check matrix $\mathcal{H} = f(\mathcal{W})$ is an array of permutation matrices. The condition that a Tanner graph is free of a given collection of subgraphs can be understood as a set of constraints imposing on such Tanner graph. Assume that the Tanner graph $G$ corresponding to $\mathcal{H}$ is required to satisfy a set of constraints. Let $\tau$ denote this set of constraints. 

The construction is based on a check and select-or-disregard procedure.  The Tanner graph of the code is built in $\rho$ stages, where $\rho$ is the row weight of $\mathcal{H}$ ($\rho$ is the number of columns of $\mathcal{W}$). Usually, $\rho$ is not pre-specified, and a code is constructed to have the rate as high as possible. Determining the exact rate is beyond the scope of this paper. At each stage, a set of $|\mathcal{Q}|$ new variable nodes are introduced that are initially not connected to the check nodes of the Tanner graph. Blocks of edges are then added to connect the new variable nodes and the check nodes. Each block of edges corresponds to a permutation matrix and hence corresponds to an element of $\mathcal{Q}$. An element of $\mathcal{Q}$ may be chosen randomly, or it may be chosen in a predetermined order. After a block of edges is tentatively added, the Tanner graph is checked for condition $\tau$. If the condition $\tau$ is violated, then that block of edges is removed and replaced by a different block. The algorithm proceeds until no block of edges can be added without violating condition $\tau$. Details of the construction is given in Algorithm \ref{ag_const}. For mathematical convenience, we append a symbol $\psi$ to the quasi group $\mathcal{Q}$ and define $f(\psi) = Z$, the all zero matrix of dimension $|\mathcal{Q}|\times |\mathcal{Q}|$. Also let $\Psi$ be a $\gamma\times 1$ matrix whose all elements are $\psi$, where $\gamma$ is the column weight of the code to be constructed.

\begin{algorithm}
\caption{Progressively Building the Tanner Graph}
\label{ag_const}
\begin{algorithmic}
\STATE $\mathcal{W}\leftarrow\gamma\times 1$ all zero matrix; $\rho\leftarrow 1$
\WHILE{$w_{\gamma,\rho}\neq\psi$}
\STATE $\mathcal{W}\leftarrow\begin{bmatrix}\mathcal{W}&\Psi\end{bmatrix}$; $S\leftarrow\mathcal{Q}$; $i\leftarrow 1$; $\rho\leftarrow\rho+1$;
\WHILE{$S\neq\emptyset~\&~i\leq\gamma$}
\STATE $w_{i,\rho}\leftarrow \xi\in S$;
\IF{$f(\mathcal{W})$ satisfies $\tau$}
\STATE $i\leftarrow i+1$;
\ENDIF
\STATE $S\leftarrow S\backslash\{\xi\}$;
\ENDWHILE
\ENDWHILE
\STATE $\rho\leftarrow\rho-1$; Delete the last column of $\mathcal{W}$;
\end{algorithmic}
\end{algorithm}

The complexity of the algorithm grows exponentially with the column weight. The speed of practical implementation of the algorithm also depends strongly on how the condition $\tau$ is checked on a Tanner graph. However, for small column weights, say 3 or 4, and small to moderate code lengths, the algorithm is well handled by state-of-the art computers. For example, with the searching techniques described in Section \ref{sect_search}, the construction of a $(1111, 808)$ code which has girth 8, minimum distance at least 10 and which contains no $(5,3)\{2\}$ trapping set takes less than 2 minutes on a 2.4 GHz computer.

\textit{Remarks:}
\begin{itemize}
\item It is worth mentioning an alternative approach in which a subgraph is described by a system of linear equations. Elements of a given matrix $\mathcal{W}$ are particular values of variables of these systems of equations. The Tanner graph corresponding to $f(\mathcal{W})$ contains the given subgraph if and only if elements of $\mathcal{W}$ form a proper solution of at least one of these linear systems of equations. For array LDPC codes, equations governing cycles and several small subgraphs have been derived in \cite{shortenedArrayCode_milenkovic} and \cite{absorbingSet_dolecek}. However, the problem of finding $\mathcal{W}$ such that its elements do not form a proper solution of any of these systems of equations is notoriously difficult.
\item The above code construction can be alternatively described as a process of progressively constructing an incidence structure. The construction begins with an incidence structure consisting of points with no lines. Blocks of parallel lines are then added based on a check and select-or disregard procedure, similar as in \cite{combinatorialLDPC_vasic} and \cite{highRateG8_vasic}.
\end{itemize}
\subsection{Determining the Collection of Forbidden Subgraphs} \label{Sect_Const}
Now we give a general rationale for deciding which trapping sets should be forbidden in the Tanner graph of a code. As previously mentioned, these trapping sets are chosen from the TSO. It is clear that if a parent trapping set is not present in a Tanner graph, then neither are its successors. Since the size of a parent trapping set is always smaller than the size of its successors, a code should be constructed so that it contains as few small parent trapping sets as possible. However, forbidding smaller trapping sets usually imposes stricter constraints on the Tanner graph, resulting in a large rate penalty. This trade off between the rate and the choice of forbidden trapping sets is also a trade off between the rate and the error floor performance. While an explicit formulation of this trade off is difficult, a good choice of forbidden trapping sets requires the analysis of decoder failures to reveal the \textit{relative harmfulness} of trapping sets. It has been pointed out that for the BSC, the slope of the frame error rate (FER) curve in the error floor region depends on the size the smallest error patterns uncorrectable by the decoder \cite{cover_milos}. We therefore introduce the notion of the relative harmfulness of trapping sets in a general setting as follows. 

Assume that under a given decoding algorithm, a code is capable of correcting any error pattern of weight $\vartheta$ but fail to correct some error patterns of weight $\vartheta+1$. If the failures of the decoders on error patterns of weight $\vartheta+1$ are due to the presence of ($a_1,b_1$) trapping sets of type $\mathcal{T}_1$ then $\mathcal{T}_1$ is the most harmful trapping set. Let us now assume that a code is constructed so that it does not contain $\mathcal{T}_1$ trapping sets and is capable of correcting any error pattern of weight $\vartheta+1$. If the presence of ($a_2,b_2$) trapping sets of type $\mathcal{T}_2$ leads to decoding failure on some error patterns of weight $\vartheta+2$ then $\mathcal{T}_2$ is the second most harmful trapping sets. The relative harmfulness of other trapping sets are also determined in this manner.

\begin{ex}
Let us consider a regular column-weight-three LDPC code of girth 8 on the BSC and assume the Gallager A/B decoding algorithm. Since such a code can correct any error pattern of weight two, we want to find subgraphs whose presence leads to decoding failure on some error patterns of weight three. Since a code can not correct three error if its Tanner graph either contain $(5,3)\{2\}$ trapping sets or contain $(8,0)\{1\}$ trapping sets, the most harmful trapping sets are the $(5,3)\{2\}$ trapping set and the $(8,0)\{1\}$ trapping set.
\end{ex}

To further explain the importance of the notion of relative harmfulness, let us slightly detour from our discussion and revisit the notion of a trapping set. A trapping set is defined as a set of variable nodes that are not eventually correct. Because trapping sets are defined in this way, it is indeed possible, in some cases, to identify some small trapping sets in a code by simulation, assuming the availability of a fast software/hardware emulator. Unfortunately, trapping sets identified in this manner generally have little significance for code construction. This is because the dynamic of an iterative decoder (except the Gallager A/B decoder on the BSC) is usually very complex and the mechanism by which the decoder fails into a trapping set is difficult to analyze and is not well understood. In many cases, the subgraphs induced by sets of non-eventually correct variable nodes are not the most harmful ones. For example, the $(6,4)$ trapping sets shown in Fig. \ref{fig_44andChild}\subref{642_lp} and \ref{fig_64andChild}\subref{641_lp} were identified in \cite{absorbingSet_dolecek} to be among the most dominant trapping sets. However, our analysis which will be presented in Section \ref{sect_codesSPA} indicates that they are not the most harmful ones. Although avoiding subgraphs induced by sets of non-eventually correct variable nodes might lead to a lower error floor, the code rate may be excessively reduced. A better solution is to increase the slope of the FER curve with the fewest possible constraints on the Tanner graph. This can only be done by avoiding the most harmful trapping sets.

Nevertheless, except for the case of the Gallager A/B algorithm on the BSC in which the relative harmfulness of a trapping set is determined by its critical number, determining the relative harmfulness of trapping sets in general is a difficult problem. The original concept of harmfulness of a trapping set can be found in early work on LDPC codes as well as importance sampling methods to analyze error floors. MacKay and Postol \cite{nearCW_mackay} were the first to discover that certain ``near codewords'' are to be blamed for  the high error floor in the Margulis code on the AWGNC. Richardson \cite{errorFloor_richarson} reproduced their results and developed a computation technique to predict the  performance of a given LDPC code in the error floor domain. He characterized the troublesome noise configurations leading to the error floor using trapping sets and described a technique (of a Monte-Carlo importance sampling type) to evaluate the error rate associated with a particular class of trapping sets. Cole \textit{et al.} \cite{06CWHG} further developed the importance sampling based method to analyze error floors of moderate length LDPC codes and we  used instantons to predict the error floors \cite{05SCCV1,04CCSV,04CCSV2}.

The main idea of our method is to determine the relative harmfulness of trapping sets from the TSO for the SPA on the BSC. It relies on the topological relationship among these trapping sets and will be presented in Section \ref{sect_codesSPA}. Before presenting this method, we describe the construction of codes for the Gallager A/B algorithm on the BSC in the next section.

\section{LDPC Codes for the Gallager A/B Algorithm on the BSC}\label{sect_GAB}

The error correction capability of regular column-weight-three LDPC codes has been studied in \cite{threeErrors_Andy, clw3paper, col3Part2} and can be summarized as follows.
\begin{itemize}
\item A column-weight-three LDPC code with Tanner graph of girth $g$ cannot correct all $g/2$ errors.
\item A column-weight-three LDPC code with Tanner graph of girth $g\geq 10$ can always correct $g/2-1$ errors.
\item A column-weight-three LDPC code with Tanner graph of girth $g=6$ can correct any two errors if and only if the Tanner graph does not contain a codeword of weight four.
\item A column-weight-three LDPC code with Tanner graph of girth $g=8$ can correct any three errors if and only if (i) the Tanner graph does not contain $(5,3)\{2\}$ trapping sets and (ii) the Tanner graph does not contain $(8,0)\{1\}$ trapping sets.
\end{itemize}

\begin{figure}
\centering
\includegraphics[width = 3.1in]{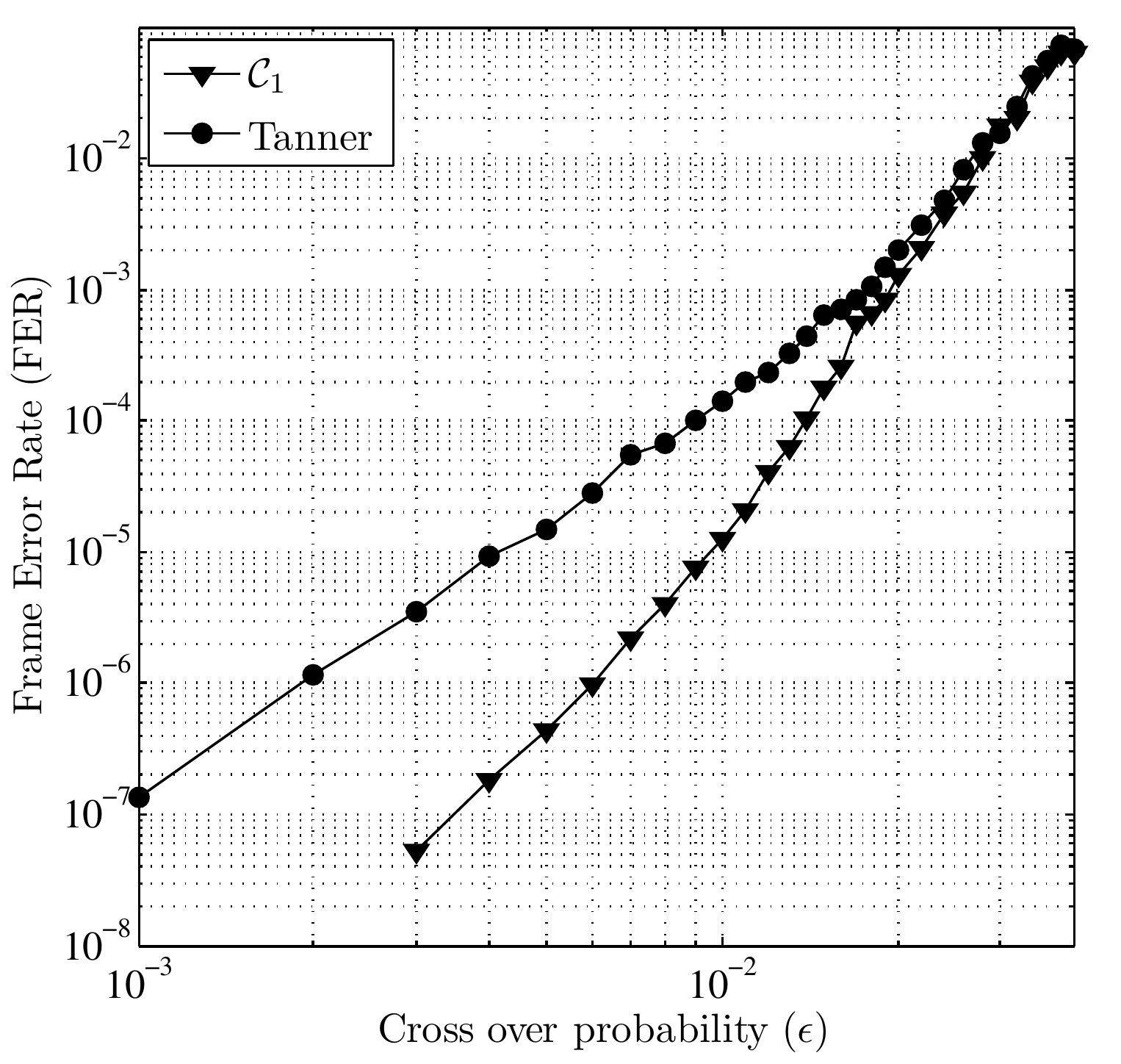}
\caption{Frame error rate performance of the Tanner code and code $\mathcal{C}_1$ under the Gallager A algorithm on the BSC with maximum 50 iterations.}
\label{fig_newTannerGA}
\end{figure}

The above conditions completely determine the set of constraints $\tau$ to be imposed on the Tanner graph of a code to achieve a given error floor performance. The necessary and sufficient conditions to correct three errors were derived in \cite{threeErrors_Andy}. These conditions require that the Tanner graph of the code has girth $g=8$, and does not contain ($5,3$) and ($8,0$) trapping set. It is obvious that the ($5,3$) is indeed the $(5,3)\{2\}$. The ($8,0$) trapping set should be understood as the $(8,0)\{1\}$ since it can be shown easily that the critical number of the $(8,0)\{2\}$ is four. In the following example, we present the construction of a code which can correct three errors.
\begin{ex}[Codes that can correct 3 errors]

The Tanner code of length 155 \cite{tannerCode_tanner} is a $(3,5)$ regular LDPC code. This code contains $(5,3)\{2\}$ trapping sets and hence can not correct three errors under the Gallager A/B algorithm on the BSC. Let $q = 31$ and $\alpha$ be a primitive element of GF($q$). Let $\mathcal{C}_1$ be an LDPC code defined by the parity check matrix $\mathcal{H}=\bar{f}(\mathcal{U}_1)$ where
\begin{equation}
\mathcal{U}_1 = \begin{bmatrix}
1&\alpha^5&\alpha^{15}&\alpha^{23}\\
\alpha^{16}&\alpha^4&\alpha^{24}&\alpha^{12}
\end{bmatrix}.\nonumber
\end{equation}

$\mathcal{C}_1$ is a $(155,64)$ LDPC code with girth $g=8$ and minimum distance $d_\mathrm{min}=12$. The Tanner graph of $\mathcal{C}_1$ contains no $(5,3)\{2\}$ trapping sets. Therefore, $\mathcal{C}_1$ is capable of correcting any three error pattern under the Gallager A/B algorithm on the BSC. The FER performance of $\mathcal{C}_1$ under the Gallager A/B algorithm over the BSC is shown in  Fig. \ref{fig_newTannerGA}. The FER performance of the Tanner code is also shown for comparison.
\end{ex}

We end this section with a remark on the harmfulness of two trapping sets with the same critical number. If two types of trapping sets $\mathcal{T}_1$ and $\mathcal{T}_2$ have the same critical number $\vartheta$, then the one with the larger number of inducing set of size $\vartheta$ is more harmful. An inducing set of a trapping set is a set of variable nodes such that if these variable nodes are initially in error then the decoder will fail on the trapping set (see \cite{ontology_vasic} for a more detailed discussion).

\section{LDPC Codes for the SPA on the BSC}\label{sect_codesSPA}
In this section, we present the construction of regular column-weight-three codes for the SPA on the BSC. The main element of the construction is the determination of the set of most harmful trapping sets. Following the discussion of the notion of relative harmfulness in Section \ref{Sect_Const}, we approach this problem as follows.

Let us consider an LDPC code $\mathcal{C}$ and assume that $\mathcal{C}$ can correct any error patterns of weight $\vartheta$ under the SPA on the BSC. We are interested in determining the trapping sets whose presence leads to decoding failure on error patterns of weight $\vartheta+1$. To simplify this problem, we only focus on initial error patterns of weight $\vartheta+1$ that surely lead to decoding failures of the Gallager A/B algorithms on the BSC. The basis for this simplification is as follows. Since it is well-known that the SPA algorithm has a superior performance in both the waterfall and error floor regions compared to that of the Gallager A/B algorithm, we surmise that an error pattern correctable by the Gallager A/B algorithm is correctable with high probability by the SPA algorithm, although this fact remains unproven. The initial error patterns of weight $\vartheta+1$ that are surely uncorrectable by the Gallager A/B algorithm can be easily derived from the TSO. 

Assume the transmission of an all zero codeword and let $\bf y$ be the received vector input to the decoder. Also assume that $\mathrm{supp}(\bf y) = \bf T_1$, a trapping set of type $\mathcal{T}_1$ from the TSO with $\vartheta+1$ variable nodes. In other words, all the $\vartheta+1$ initially corrupt variable nodes belong to the trapping set $\bf T_1$. This error pattern results in a decoding failure of the Gallager A/B algorithm and  hence is an initial error pattern of interest. As the decoder operates by passing messages along edges of the Tanner graph, the decoding outcome depends heavily on the immediate neighborhood of the subgraph induced by variable nodes in $\bf T_1$. In many cases, a decoding failure will occur if $\bf T_1$ generates a trapping set $\bf T_2$ of type $\mathcal{T}_2$, where $\mathcal{T}_2$ is a successor of $\mathcal{T}_1$. In such cases, the presence of $\bf T_2$ in a code make it incapable of correcting any error pattern of size $\vartheta+1$ and hence $\bf T_2$ is a harmful trapping set.

To evaluate the harmfulness of the $\mathcal{T}_2$ trapping set, all initial error patterns that consist of variable nodes of a $\mathcal{T}_1$ trapping set must be considered. Let $\mathscr{T}$ be the set of all trapping sets of type $\mathcal{T}_1$. Partition $\mathscr{T}$ into two disjoint sets $\mathscr{T}_1$ and $\mathscr{T}_2$ such that a trapping set in $\mathscr{T}_1$ generates at least one $\mathcal{T}_2$ trapping set while a trapping set in $\mathscr{T}_2$ does not generate any $\mathcal{T}_2$ trapping set. For each trapping set $\bf T_i \in \mathscr{T}$, perform decoding on the input vector $\bf y_i$ where $\mathrm{supp}(\bf y_i) = \bf T_i$, at a cross over probability $\epsilon$ of the channel. Let $\mathscr{T}_\mathrm{s}$ be the set of trapping sets $\bf T_i \in \mathscr{T}$ such that decoding is successful upon error pattern $\bf y_i$. Define $\chi_1(\epsilon)$ and $\chi_2(\epsilon)$, the rate of successful decoding for trapping sets in $\mathscr{T}_1$ and $\mathscr{T}_2$ at the cross over probability $\epsilon$ of the channel, as follows
\begin{eqnarray}
\chi_1(\epsilon) = \frac{|\mathscr{T}_1\cap\mathscr{T}_\mathrm{s}|}{|\mathscr{T}_1|}\\
\chi_2(\epsilon) = \frac{|\mathscr{T}_2\cap\mathscr{T}_\mathrm{s}|}{|\mathscr{T}_2|}
\end{eqnarray}

The harmfulness of $\mathcal{T}_2$ trapping sets of $\mathcal{C}$ is evaluated by comparing $\chi_1(\epsilon)$ and $\chi_2(\epsilon)$ for a wide range of $\epsilon$. The larger the difference $\chi_2(\epsilon)-\chi_1(\epsilon)$, the more harmful $\mathcal{T}_2$ trapping sets are. The harmfulness of $\mathcal{T}_2$ trapping sets is also compared with the harmfulness of other successor trapping sets of $\mathcal{T}_1$, which is determined in the same fashion. Once the most harmful trapping sets have been determined, a code is constructed so that it does not contain these trapping sets.

We note that this characterization of relative harmfulness, although heuristic, plays a critical role in the construction of good high rate codes as no explicit quantification of harmfulness of trapping sets is known. This characterization of harmfulness also helps a code designer to determine more or less the exact subgraphs that are responsible for a certain type of decoding failure. It is therefore superior to searching for trapping sets by simulation.

\begin{figure}
\centering
\includegraphics[width = 3.1in]{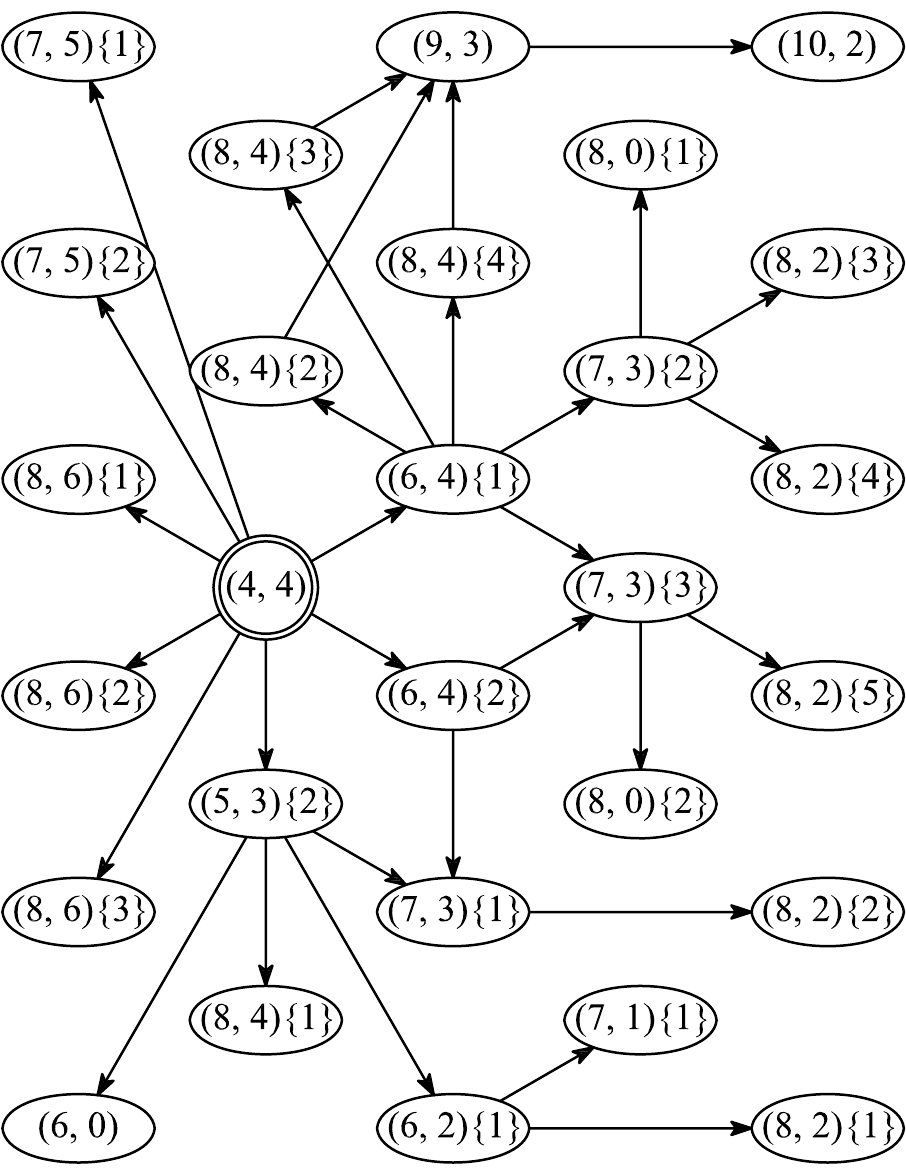}
\caption{Hierarchy of trapping sets originating from the $(4,4)$ trapping set for regular column-weight-three codes of girth 8.}
\label{fig_tree}
\end{figure}

We continue our discussion with three case studies in which we evaluate (i) the relative harmfulness of the $(6,2)\{1\}$ and $(8,2)$ trapping sets, (ii) the relative harmfulness the $(5,3)\{2\}$ trapping set and (iii) the relative harmfulness of the $(7,3)$, $(9,3)$ and $(10,2)$ trapping sets. For a better illustration of the relationship among these trapping sets, a hierarchy of trapping sets originating from the $(4,4)$ trapping set is shown in Fig. \ref{fig_tree}. For the first case, we present a detailed analysis. For the other two cases, we only give the results of the analysis. The analysis to be presented is a step towards the guaranteed correction of four, five and six errors under the SPA on the BSC. For simplicity, we assume that codes have girth $g=8$ in all examples, although the method of construction can be applied to girth 6 codes to likely result in higher rate codes.

\subsection{The Harmfulness of the $(6,2)\{1\}$ and $(8,2)$ Trapping Sets} 
Since we consider codes with girth $g=8$, let us start with an existing code of such girth. Consider the $(530,373)$ integer lattice code (or shortened array code \cite{shortenedArrayCode_milenkovic}) given in \cite{highRateG8_vasic}. This code has minimum distance $d_{min} = 8$ and hence is unable to correct all weight-four error patterns. Clearly, the first step towards the guaranteed correction of four errors is to eliminate the $(6,0)\{1\}$, $(8,0)\{1\}$ and the $(8,0)\{2\}$ trapping sets, which are the low weight codewords. We therefore construct a code with minimum distance $d_{min} \geq 10$. Let $q=53$ and $\alpha$ be a primitive element of GF($q$) and let $\tau$ specify that the Tanner graph of a code has girth $g=8$ and contain no $(6,0)\{1\}$, $(8,0)\{1\}$ and $(8,0)\{2\}$ trapping sets. Using the method of construction described in Section \ref{subsect_construct}, we obtain a regular column-weight-three code $\mathcal{C}_2$ with parity check matrix $\mathcal{H}_2 = \bar{f}(\mathcal{U}_2)$ where 
\begin{equation}
\mathcal{U}_2=\begin{bmatrix}
1&\alpha^2&\alpha^4&\alpha^6&\alpha^7&\alpha^{11}&\alpha^{12}&\alpha^{14}&\alpha^{27}\\
\alpha&\alpha^3&\alpha^5&\alpha^8&\alpha^{10}&\alpha^{13}&\alpha^9&\alpha^{38}&\alpha^{51}
\end{bmatrix}.\nonumber
\end{equation}

$\mathcal{C}_2$ is a $(530,373)$ code. Similar to the above integer lattice code, $\mathcal{C}_2$ has column weight 3, row weight 10 and rate $R = 0.7$. 

The Tanner graph of $\mathcal{C}_2$ contains 17066 $(4,4)$ trapping sets. We partition the collection of $(4,4)$ trapping sets into nine disjoint sets $\mathscr{T}_1, \mathscr{T}_2,\ldots,\mathscr{T}_9$ based on whether a $(4,4)$ trapping set generate $(5,3)\{2\}$, $(6,2)\{1\}$, $(8,2)$ or $(10,0)$ trapping sets. Note that, for simplicity, we do not differentiate among different $(8,2)$ and $(10,0)$ trapping sets in this analysis, although a more detailed treatment may reveal some difference in the harmfulness of those trapping sets. The classification and sizes of different sets of $(4,4)$ trapping sets are shown in Table \ref{tb_type44530New}. 
\begin{table}[htb]
\caption{Disjoint Sets of $(4,4)$ trapping sets in the LDPC code $\mathcal{C}_2$. A \checkmark indicates that the $(4,4)$ trapping sets in $\mathscr{T}_i$ generate at least one corresponding trapping set.}
\begin{center}
\begin{tabular}{|c||c|c|c|c|c|}\hline
\multirow{2}{*}{Sets $\mathscr{T}_i$}  &\multicolumn{4}{c|}{Trapping Sets Generated by $\mathscr{T}_i$}&\multirow{2}{*}{Total}\\\cline{2-5}
 						 &$(5,3)\{2\}$     & $(6,2)\{1\}$&$(8,2)$      & $(10,0)$ & \\\hline\hline
$\mathscr{T}_1$				  	 &			           &				     &             &           &4982				\\\hline
$\mathscr{T}_2$				  	 &			           &				     &             &\checkmark &53					\\\hline
$\mathscr{T}_3$				  	 &\checkmark       &				     &             & 					 &424				\\\hline
$\mathscr{T}_4$				  	 &					       &				     &\checkmark   & 					 &7314				\\\hline
$\mathscr{T}_5$				  	 &					       &				     &\checkmark   &\checkmark &371				\\\hline
$\mathscr{T}_6$				  	 &\checkmark       &				     &\checkmark   & 					 &1855				\\\hline
$\mathscr{T}_7$				  	 &\checkmark       &				     &\checkmark   &\checkmark &106				\\\hline
$\mathscr{T}_8$				  	 &\checkmark			 &\checkmark   &             & 					 &1007				\\\hline
$\mathscr{T}_9$				  	 &\checkmark			 &\checkmark   &\checkmark   & 					 &954				\\\hline
\multicolumn{5}{|c|}{\textbf{Total}}&\textbf{17066}\\\hline
\end{tabular}
\end{center}
\label{tb_type44530New}
\end{table}

\begin{figure*}
\centering
\includegraphics[]{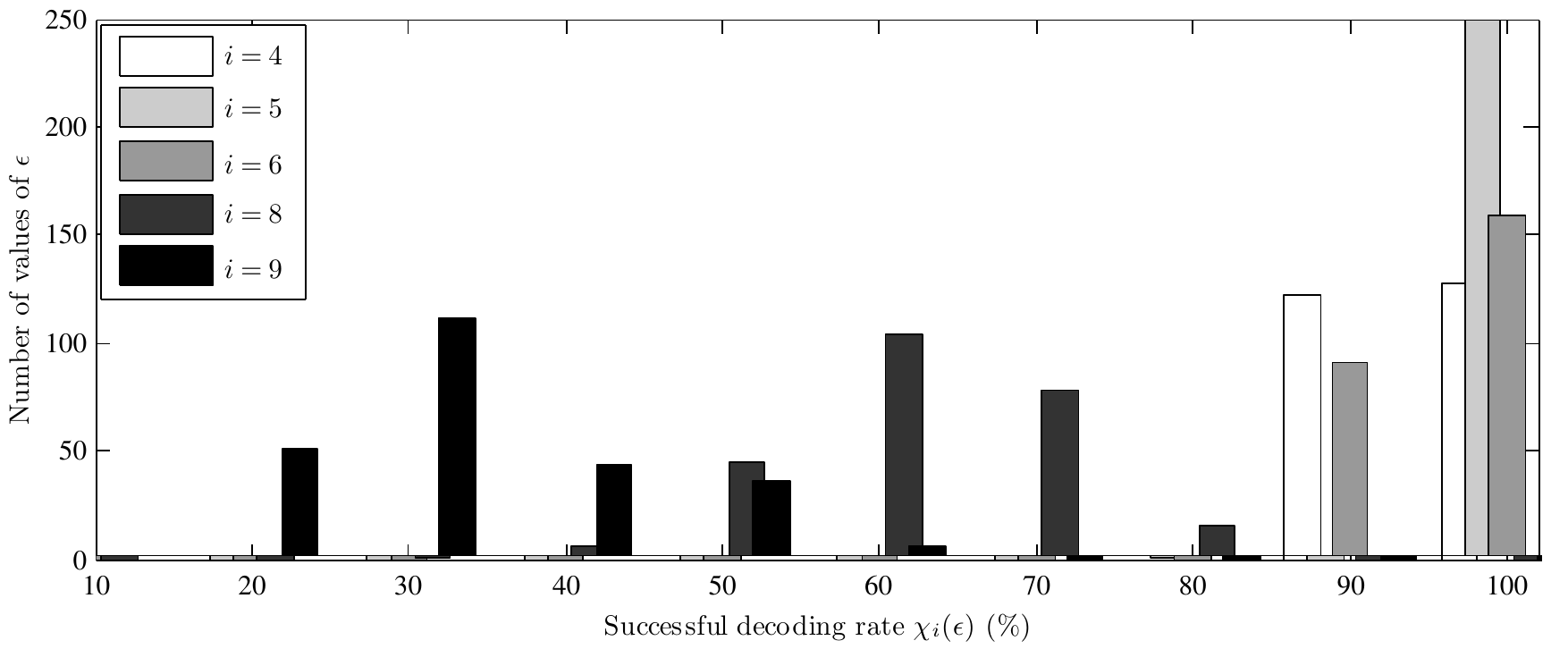}
\caption{The rate of successful decoding for different sets of $(4,4)$ trapping sets in code $\mathcal{C}_2$}
\label{fig_44Code530New}
\end{figure*}

To evaluate the harmfulness of the $(5,3)\{2\}$, $(6,2)\{1\}$, $(8,2)$ and $(10,0)$ trapping sets, we perform decoding on all input vectors $\bf y_i$ where $\mathrm{supp}(\bf y_i) = \bf T_i$, a $(4,4)$ trapping set of $\mathcal{C}_2$. The result is as follows. For the trapping sets in $\mathscr{T}_1$, $\mathscr{T}_1$, $\mathscr{T}_2$ and $\mathscr{T}_7$, the decoder successfully decodes all input vectors $\bf y_i$ at all the 250 values of $\epsilon$ that have been considered, i.e. $\chi_1(\epsilon) = \chi_2(\epsilon) = \chi_3(\epsilon) = \chi_7(\epsilon) = 100\%$ $\forall \epsilon$. For the trapping sets in $\mathscr{T}_4$, $\mathscr{T}_5$, $\mathscr{T}_6$, $\mathscr{T}_8$ and $\mathscr{T}_9$, the rate of successful decoding is shown in form of histogram in Fig. \ref{fig_44Code530New}. As an example of how to interpret the result, consider the trapping sets in $\mathscr{T}_6$. It can be seen that there are about 160 values (65\%) of $\epsilon$ at which decoding is succesful for all input vectors $\bf y_i$. For about 90 values (30\%) of $\epsilon$, decoding is succesful for approximately nine out of ten input vectors $\bf y_i$.

The following facts can be observed:
\begin{itemize}
\item The $(4,4)$ trapping sets in $\mathscr{T}_1,\mathscr{T}_2$ and $\mathscr{T}_3$ do not generate either $(6,2)\{1\}$ or $(8,2)$ trapping sets. The rate of successful decoding is 100\% for all tested values of $\epsilon$.
\item The $(4,4)$ trapping sets in $\mathscr{T}_4, \mathscr{T}_5$ and $\mathscr{T}_6$ generate at least one $(8,2)$ trapping set. Decoding is not always successful, but the rate of successful decoding is more than 90\% for all tested values of $\epsilon$.
\item The $(4,4)$ trapping sets in $\mathscr{T}_8$ generate at least one $(6,2)\{1\}$ trapping set. The rate of successful decoding $\chi_8(\epsilon)$ is significantly lower in general compared to type $\chi_4(\epsilon),\chi_5(\epsilon)$ and $\chi_6(\epsilon)$.
\item The $(4,4)$ trapping sets in $\mathscr{T}_9$ generate at least one $(6,2)\{1\}$ and one $(8,2)$ trapping set. The rate of successful decoding $\chi_9(\epsilon)$ is lowest in general. 
\item The $(4,4)$ trapping sets in $\mathscr{T}_6$ generate at least one $(5,3)\{2\}$ trapping set while the ones in $\mathscr{T}_4$ do not. In general, $\chi_6(\epsilon)<\chi_4(\epsilon)$.
\item The $(4,4)$ trapping sets in $\mathscr{T}_5$ generate at least one $(10,0)$ trapping set while the ones in $\mathscr{T}_4$ do not. In general, $\chi_5(\epsilon)>\chi_4(\epsilon)$. 
\item The $(4,4)$ trapping sets in $\mathscr{T}_7$ generate at least one $(10,0)$ trapping set while the ones in $\mathscr{T}_6$ do not. $\chi_7(\epsilon) = 100\%$ for all tested values of $\epsilon$.
\end{itemize}

The above observations strongly suggest that both $(6,2)\{1\}$ and $(8,2)$ trapping sets are harmful. However, the harmfulness of the $(6,2)\{1\}$ trapping set is much more evident than the harmfulness of the $(8,2)$ trapping set. Besides, it is interesting to notice that $\chi_7(\epsilon) = 100\%$ for all tested values of $\epsilon$. All $(4,4)$ trapping sets in $\mathscr{T}_7$ generate at least one $(8,2)$ trapping set, one $(5,3)\{2\}$ trapping set and one ($10,0$) trapping set. In this case, the presence of ($10,0$) trapping sets seem to ``help'' decoding. This ``positive'' effect of ($10,0$) trapping sets can also be seen when comparing $\chi_5(\epsilon)$ and $\chi_4(\epsilon)$. Finally, by comparing $\chi_6(\epsilon)$ and $\chi_4(\epsilon)$, it is suggestive that the $(5,3)\{2\}$ trapping sets have some negative effect on decoding if the $(4,4)$ trapping set generate $(8,2)$ and ($10,0$) trapping sets.

To further verify our prediction on the harmfulness of the $(6,2)\{1\}$ and $(8,2)$ trapping sets, we construct another code with the same parameters as those of $\mathcal{C}_2$. We denote this code by $\mathcal{C}_3$. The Tanner graph of $\mathcal{C}_3$ has stronger constraints than the Tanner graph of $\mathcal{C}_2$ as it contains neither $(6,2)\{1\}$ nor ($10,0$) trapping sets. Since ($10,0$) trapping sets are not presented, $\mathcal{C}_3$ has minimum distance at least 12.

Let $\mathcal{C}_3$ be defined by the parity check matrix $\mathcal{H}_3 = \bar{f}(\mathcal{U}_3)$ where
\begin{equation}
\mathcal{U}_3=
\begin{bmatrix}
1&\alpha^2&\alpha^4&\alpha^{15}&\alpha^{17}&\alpha^{26}&\alpha^{31}&\alpha^{33}&\alpha^{36}\\
\alpha^{30}&\alpha^{16}&\alpha&\alpha^{19}&\alpha^7&\alpha^{34}&\alpha^3&\alpha^{8}&\alpha^{22}
\end{bmatrix}.\nonumber
\end{equation}
The Tanner graph of $\mathcal{C}_3$ contains 16483 $(4,4)$ trapping sets, which can be partitioned into four disjoint sets as shown in Table \ref{tb_type44530No62}.
\begin{table}[htb]
\caption{Types of $(4,4)$ trapping sets in the $(530,373)$ LDPC code $\mathcal{C}_3$}
\begin{center}
\begin{tabular}{|c||c|c|c|}\hline
\multirow{2}{*}{Sets $\mathscr{T}_i$}  &\multicolumn{2}{c|}{Trapping Sets Generated by $\mathscr{T}_i$}&\multirow{2}{*}{Total}\\\cline{2-3}
 						               &~~~$(5,3)\{2\}$~~~     &$(8,2)$&    \\\hline\hline
$\mathscr{T}_1$				  	 &			           &				    &6890\\\hline
$\mathscr{T}_2$				  	 &\checkmark       &				   	&795 \\\hline
$\mathscr{T}_3$				  	 &    				     &\checkmark  &6890 \\\hline
$\mathscr{T}_4$				  	 &\checkmark			 &\checkmark  &1908 \\\hline
\multicolumn{3}{|c|}{\textbf{Total}}&\textbf{16483}\\\hline
\end{tabular}
\end{center}
\label{tb_type44530No62}
\end{table}

\begin{figure}
\centering
\includegraphics[]{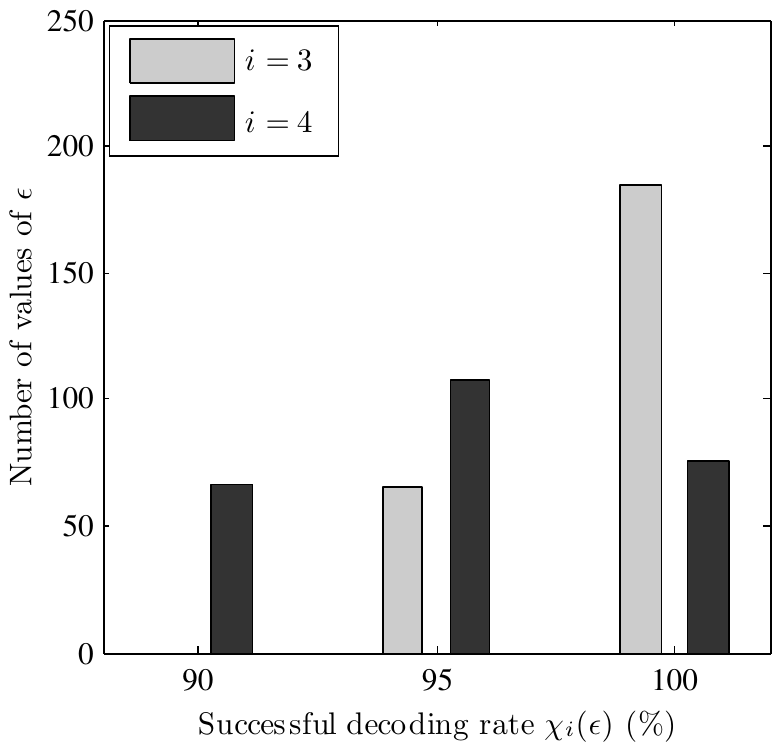}
\caption{The rate of successful decoding for different sets of $(4,4)$ trapping sets in code $\mathcal{C}_3$}
\label{fig_44Code530No62}
\end{figure}

We again perform decoding on all input vectors $\bf y_i$ where $\mathrm{supp}(\bf y_i) = \bf T_i$, a $(4,4)$ trapping set of $\mathcal{C}_3$. The rate of successful decoding for trapping sets in $\mathscr{T}_3$ and $\mathscr{T}_4$ is shown in form of histogram in Fig. \ref{fig_44Code530No62}. For trapping sets in $\mathscr{T}_1$ and $\mathscr{T}_2$, decoding is always successful.

It can be seen that the results are consistent with the previously obtained results. Decoding is always successful for the $(4,4)$ trapping sets which generate neither $(6,2)\{1\}$ nor $(8,2)$ trapping sets. Besides, $\chi_3(\epsilon)>\chi_4(\epsilon)$ in general since the $(4,4)$ trapping sets in $\mathscr{T}_3$ do not generate $(5,3)\{2\}$ trapping sets. These results validate our prediction on the harmfulness of successors of the $(4,4)$ trapping set. We have repeated the experiment for a collection of codes whose Tanner graphs do not contain either $(6,2)\{1\}$ or $(8,2)$ trapping sets. The consistency of the results led us to the following conjecture.
\begin{conj} A regular column-weight-three code of girth $g=8$ can correct any error pattern of weight 4 consisting of variable nodes of an eight cycle under the SPA on the BSC if its Tanner graph contain neither $(6,2)$ nor $(8,2)$ trapping sets.
\end{conj}

We remark that this conjecture only gives a sufficient condition. A code may correct any error pattern of weight 4 even if its Tanner graph contains $(8,2)\{2\}$ trapping sets. For example, consider the Tanner code of length 155. The Tanner graph of this code does not contain $(6,2)\{1\}$ trapping set, but it contains $(8,2)\{2\}$ trapping sets. However, decoding is always successful for all the $(4,4)$ trapping sets at any value of $\epsilon$. It might be possible to find a better sufficient condition by taking into account bigger trapping sets, but such analysis appears to be difficult.

\begin{ex}\label{ex_530}
The FER performance of $\mathcal{C}_2$, $\mathcal{C}_3$ and the $(530,373)$ integer lattice code under the SPA with 100 iterations on the BSC is shown in  Fig. \ref{fig_fer530}. For comparison,  Fig. \ref{fig_fer530} also shows the FER performance of a $(530,373)$ LDPC code constructed using the PEG algorithm \cite{peg_hu}. This PEG code has girth $g=6$ and minimum distance $d_{min} = 6$. Clearly, $\mathcal{C}_3$ whose Tanner graph is free $(6,2)$ trapping sets, has the best performance. Although the Tanner graph of $\mathcal{C}_2$ contains some $(8,2)$ trapping sets, it still outperforms the PEG code. The integer lattice code has the worst performance although it has girth $g = 8$.
\begin{figure}
\centering
\includegraphics[width = 3.1in]{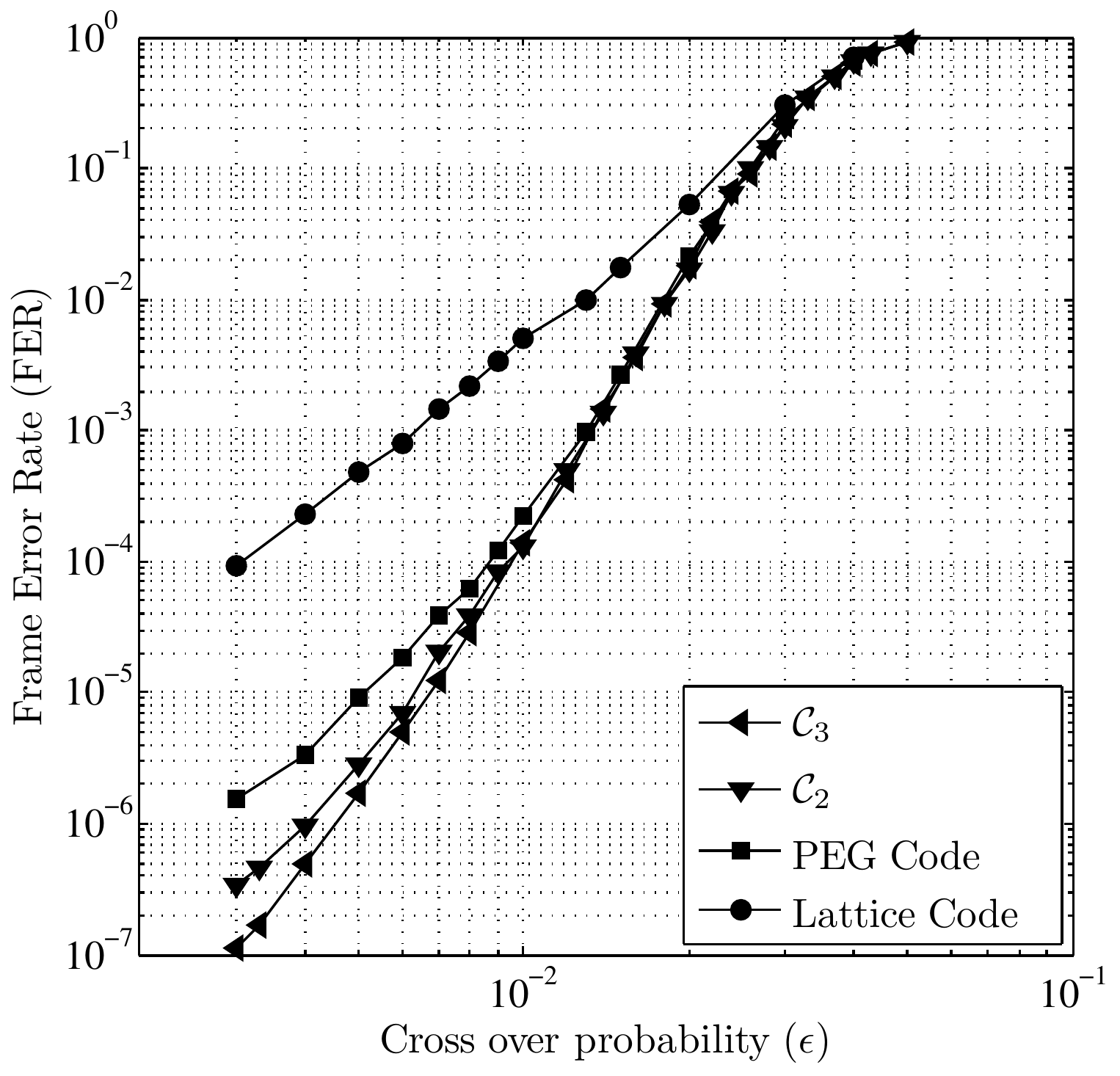}
\caption{Frame error rate performance of codes in Example \ref{ex_530} under the SPA on the BSC.}
\label{fig_fer530}
\end{figure}
\end{ex}
\begin{ex}\label{ex_810}
Let $q = 3^4$ and let $\mathcal{C}_4$ be defined by the parity check matrix $\mathcal{H}_4 = \bar{f}(\mathcal{U}_5)$ where
\begin{equation}
\mathcal{U}_4=\\
\begin{bmatrix}
\alpha^2&\alpha^6&\alpha^9&\alpha^{31}&\alpha^{33}&\alpha^{39}&\alpha^{57}&\alpha^{60}&\alpha^{67}\\
\alpha^{55}&\alpha^{12}&\alpha^{28}&\alpha^{46}&\alpha^{78}&\alpha^{37}&\alpha^{61}&\alpha^{76}&\alpha^{44}
\end{bmatrix}.\nonumber
\end{equation}

$\mathcal{C}_4$ is a $(810,569)$ code with column weight 3, row weight 10 and rate $R = 0.7$. The Tanner graph of $\mathcal{C}_4$ has girth $g=8$ and does not contain either $(6,2)$ or $(8,2)$ trapping sets. The FER performance of $\mathcal{C}_4$  under the SPA with 100 iterations on the BSC is shown in  Fig. \ref{fig_fer810}. For comparison,  Fig. \ref{fig_fer810} also shows the FER performance of a $(810,567)$ PEG constructed code. This code has girth $g = 8$. It can be seen that $\mathcal{C}_4$ has a lower floor than the PEG code.
\begin{figure}
\centering
\includegraphics[width = 3.1in]{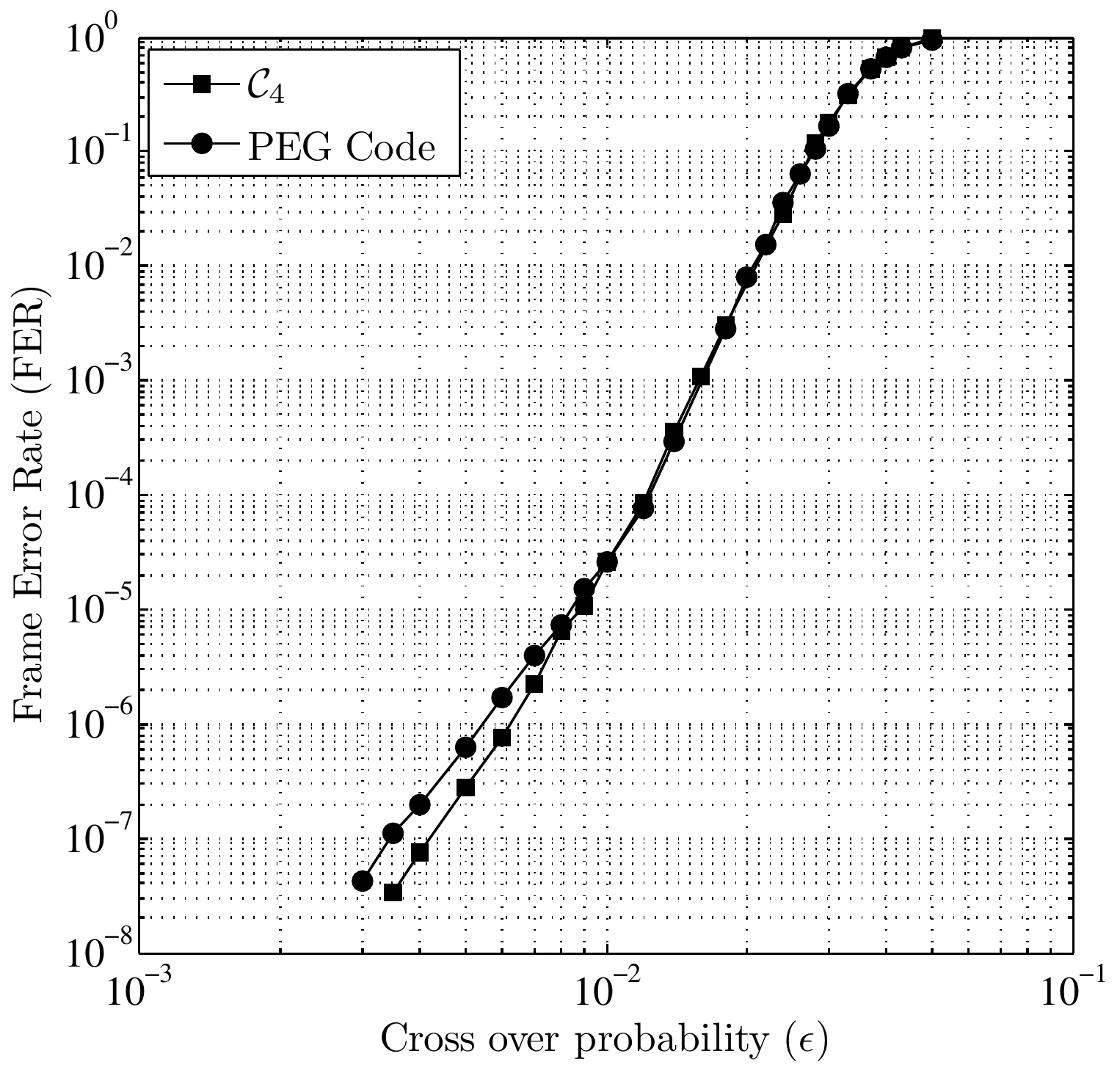}
\caption{Frame error rate performance of codes in Example \ref{ex_810} under the SPA on the BSC.}
\label{fig_fer810}
\end{figure}
\end{ex}

\subsection{The Harmfulness the $(5,3)\{2\}$ Trapping Set}
Assuming the guaranteed correction of four errors, we are now interested in finding trapping sets whose presence leads to decoding failure on some error patterns of weight five. There are two trapping sets with five variable nodes from the TSO that can be present in the Tanner graph of a regular column-weight-three LDPC code with girth $g=8$: the $(5,3)\{2\}$ trapping set and the $(5,5)$ trapping set. The result of our analysis indicates that $(5,3)\{2\}$ trapping sets are the most harmful and should be forbidden in the Tanner graph of a code.

\begin{ex}\label{ex_no53}
Let $q = 211$ and let $\mathcal{C}_5$ be defined by the parity check matrix $\mathcal{H}_5 = \bar{f}(\mathcal{U}_5)$ where $\mathcal{U}_5$ is shown in (\ref{eqn_dbl_x}). $\mathcal{C}_5$ is a $(3165,2554)$ code with column weight 3, row weight 15 and rate $R = 0.8$. The Tanner graph of $\mathcal{C}_5$ has girth $g=8$ and does not contain either $(5,3)\{2\}$ or $(8,2)$ trapping sets. The FER performance of $\mathcal{C}_5$  under the SPA with 100 iterations on the BSC is shown in  Fig. \ref{fig_ne53fer}. For comparison,  Fig. \ref{fig_ne53fer} also shows the FER performance of a ($3150, 2520)$ regular QC LDPC code constructed using array masking proposed in \cite{qcFiniteField_lanLin}. The parity check matrix of this code is a $10 \times 50$ array of $63 \times 63$ circulants or zero matrices, which has column weight 3 and row weight 15. This code has girth $g = 8$. It can be seen that $\mathcal{C}_4$ has a lower error floor than the code constructed using array masking.

\begin{figure*}[!b]
\vspace*{4pt}
\hrulefill
\normalsize
\begin{equation}
\label{eqn_dbl_x}
\mathcal{U}_5=\left[
\begin{array}{ccccccccccccccc}
\alpha^{8}&\alpha^{23}&\alpha^{47}&\alpha^{54}&\alpha^{55}&\alpha^{67}&\alpha^{78}&\alpha^{90}&\alpha^{108}&\alpha^{169}&\alpha^{177}&\alpha^{187}&\alpha^{192}&\alpha^{193}\\
\alpha^{61}&\alpha^{189}&\alpha^{190}&\alpha^{171}&\alpha^{170}&\alpha^{132}&\alpha^{182}&\alpha^{128}&\alpha^{71}&\alpha^{117}&\alpha^{129}&\alpha^{10}&\alpha^{160}&\alpha^{64}
\end{array}\right].
\end{equation}
\begin{equation}
\label{eqn_dbl_x2}
\mathcal{U}_6=\left[
\begin{array}{cccccccccccc}
\alpha^{12}&\alpha^{15}&\alpha^{28}&\alpha^{34}&\alpha^{37}&\alpha^{57}&\alpha^{75}&\alpha^{85}&\alpha^{111}&\alpha^{157}&\alpha^{166}\\
\alpha^{163}&\alpha^{175}&\alpha^{60}&\alpha^{25}&\alpha^{118}&\alpha^{167}&\alpha^{156}&\alpha^{142}&\alpha^{30}&\alpha^{155}&\alpha^{31}
\end{array}\right].
\end{equation}
\begin{equation}
\label{eqn_dbl_x3}
\mathcal{U}_7=\left[
\begin{array}{ccccccccccccc}
\alpha^{18}&\alpha^{88}&\alpha^{112}&\alpha^{142}&\alpha^{157}&\alpha^{186}&\alpha^{196}&\alpha^{197}&\alpha^{228}&\alpha^{246}&\alpha^{288}&\alpha^{316}\\
\alpha^{25}&\alpha^{73}&\alpha^{155}&\alpha^{287}&\alpha^{328}&\alpha^{151}&\alpha^{75}&\alpha^{324}&\alpha^{148}&\alpha^{248}&\alpha^{62}&\alpha^{70}
\end{array}\right].
\end{equation}
\end{figure*}
\begin{figure}
\centering
\includegraphics[width = 3.1 in]{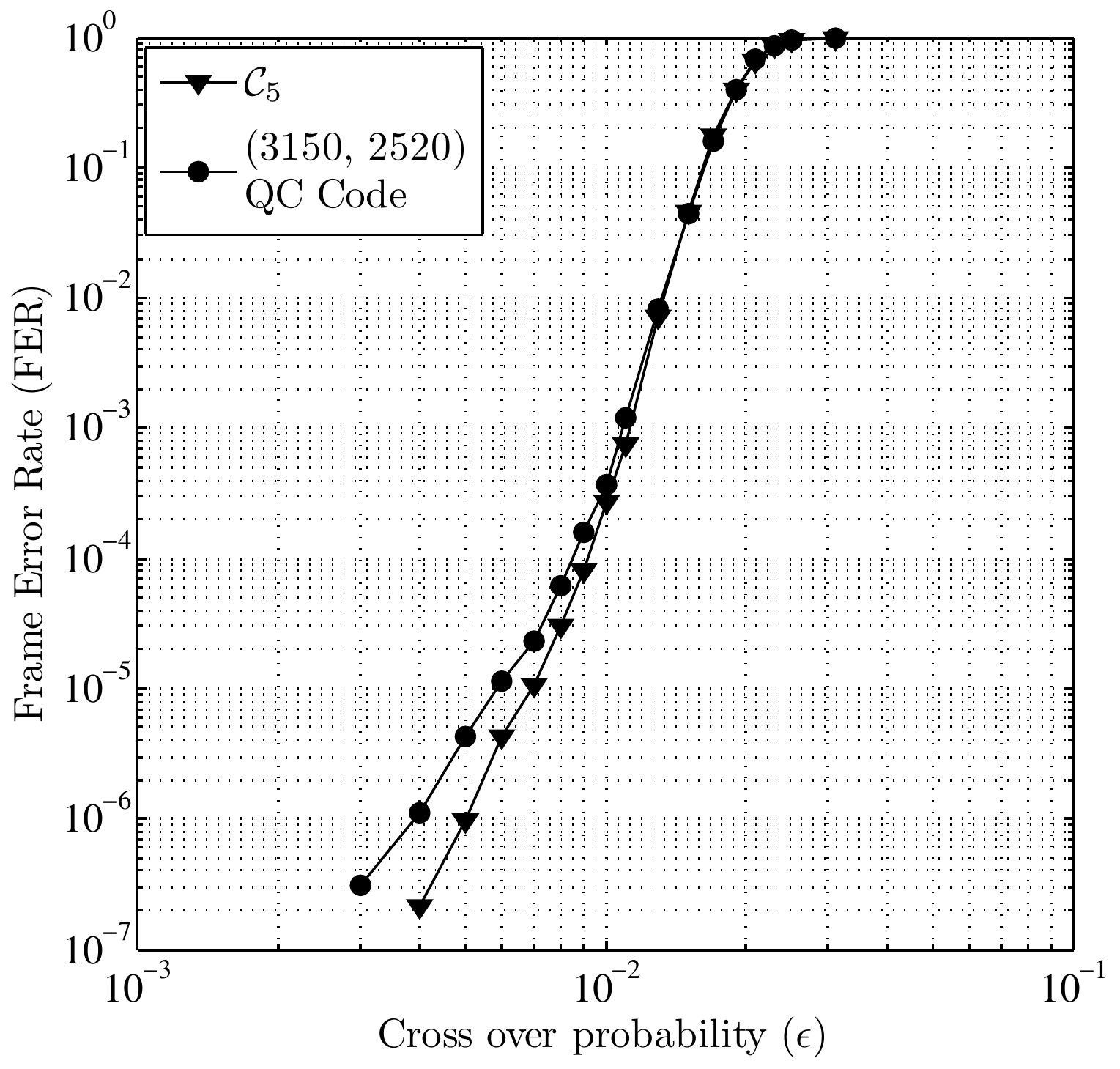}
\caption{Frame error rate performance of codes in Example \ref{ex_no53} under the SPA on the BSC.}
\label{fig_ne53fer}
\end{figure}
\end{ex}

\subsection{The Harmfulness of $(7,3)$, $(9,3)$ and $(10,2)$ Trapping Sets}
With the above results, we now consider codes free of $(5,3)\{2\}$ and $(8,2)$ trapping sets and aim for the guaranteed correction of six errors. To guarantee the correction of six errors, codes must have minimum distances $d_{min}\geq 14$. In other words, their Tanner graphs should be free of $(a,0)$ trapping sets $\forall a\leq 12$. Similar to the previous discussions, we analyze error patterns of weight six, focusing on those consisting of variable nodes of a trapping set. There are three trapping sets of size 6 from the TSO that can be present in the Tanner graph of a regular column-weight-three LDPC code with girth $g=8$: the $(6,6)$ trapping set, $(6,4)\{1\}$ trapping set and the $(6,4)\{2\}$ trapping set. The results of our analysis and experiments suggest that the following trapping sets are harmful (in a decreased order of harmfulness):
\begin{itemize}
\item The $(7,3)\{2\}$ trapping set and the $(7,3)\{3\}$ trapping set.
\item The $(10,2)$ trapping sets which are successors of the $(9,3)$ trapping sets below.
\item The $(9,3)$ trapping sets which are successors of the $(8,4)\{2\}$, $(8,4)\{3\}$ and $(8,4)\{4\}$ trapping sets (see Fig. \ref{fig_tree} for an illustration of the relationship among these trapping sets).
\end{itemize}

\begin{ex}\label{ex_no73102}
Let $q = 199$ and let $\mathcal{C}_6$ be defined by the parity check matrix $\mathcal{H}_6 = \bar{f}(\mathcal{U}_6)$ where $\mathcal{U}_6$ is shown in (\ref{eqn_dbl_x2}). $\mathcal{C}_6$ is a $(2388,1793)$ code with column weight 3, row weight 12 and rate $R = 0.75$. The Tanner graph of $\mathcal{C}_6$ has girth $g=8$ and does not contain either $(5,3)\{2\}$ or $(7,3)$ trapping sets and neither does it contain $(10,2)$ trapping sets that are generated by either $(8,4)\{2\}$, $(8,4)\{3\}$ or $(8,4)\{4\}$ trapping sets. The FER performance of $\mathcal{C}_6$  under the SPA with 100 iterations on the BSC is shown in  Fig. \ref{fig_ne73102fer}. For comparison,  Fig. \ref{fig_ne73102fer} also shows the FER performance of a ($3150, 2518)$ PEG code. This code has girth $g = 8$. It can be seen that $\mathcal{C}_6$ has a lower error floor than the PEG code.

\begin{figure}
\centering
\includegraphics[width = 3.1 in]{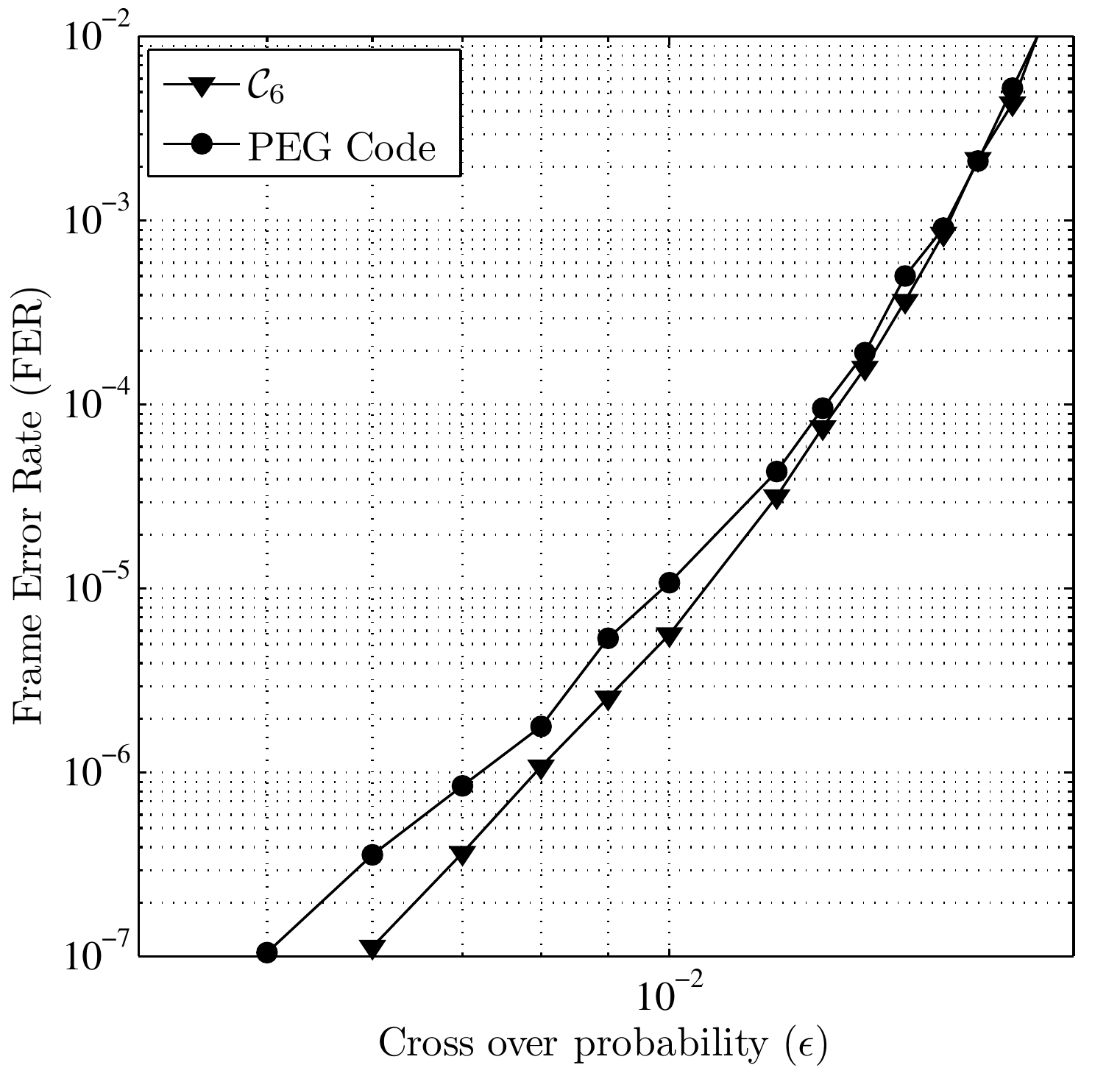}
\caption{Frame error rate performance of codes in Example \ref{ex_no73102} under the SPA on the BSC.}
\label{fig_ne73102fer}
\end{figure}
\end{ex}   
\begin{ex}\label{ex_no7393}
Let $q = 337$ and let $\mathcal{C}_7$ be defined by a parity check matrix $\mathcal{H}_7 = \bar{f}(\mathcal{U}_7)$ where $\mathcal{U}_7$ is shown in (\ref{eqn_dbl_x3}). $\mathcal{C}_7$ is a $(4381,3372)$ code with column weight 3, row weight 13 and rate $R = 0.77$. The Tanner graph of $\mathcal{C}_7$ has girth $g=8$ and does contain either $(5,3)\{2\}$ or $(7,3)$ trapping sets and neither does it contain $(9,3)$ trapping sets that are generated by either $(8,4)\{2\}$, $(8,4)\{3\}$ or $(8,4)\{4\}$ trapping sets. The FER performance of $\mathcal{C}_7$  under the SPA with 100 iterations on the BSC is shown in  Fig. \ref{fig_ne7393fer}. For comparison,  Fig. \ref{fig_ne7393fer} also shows the FER performance of a $(4381,3370)$ PEG code. This code has girth $g = 8$. It can be seen that $\mathcal{C}_7$ has a lower error floor than that of the PEG code.  
\begin{figure}
\centering
\includegraphics[width = 3.1 in]{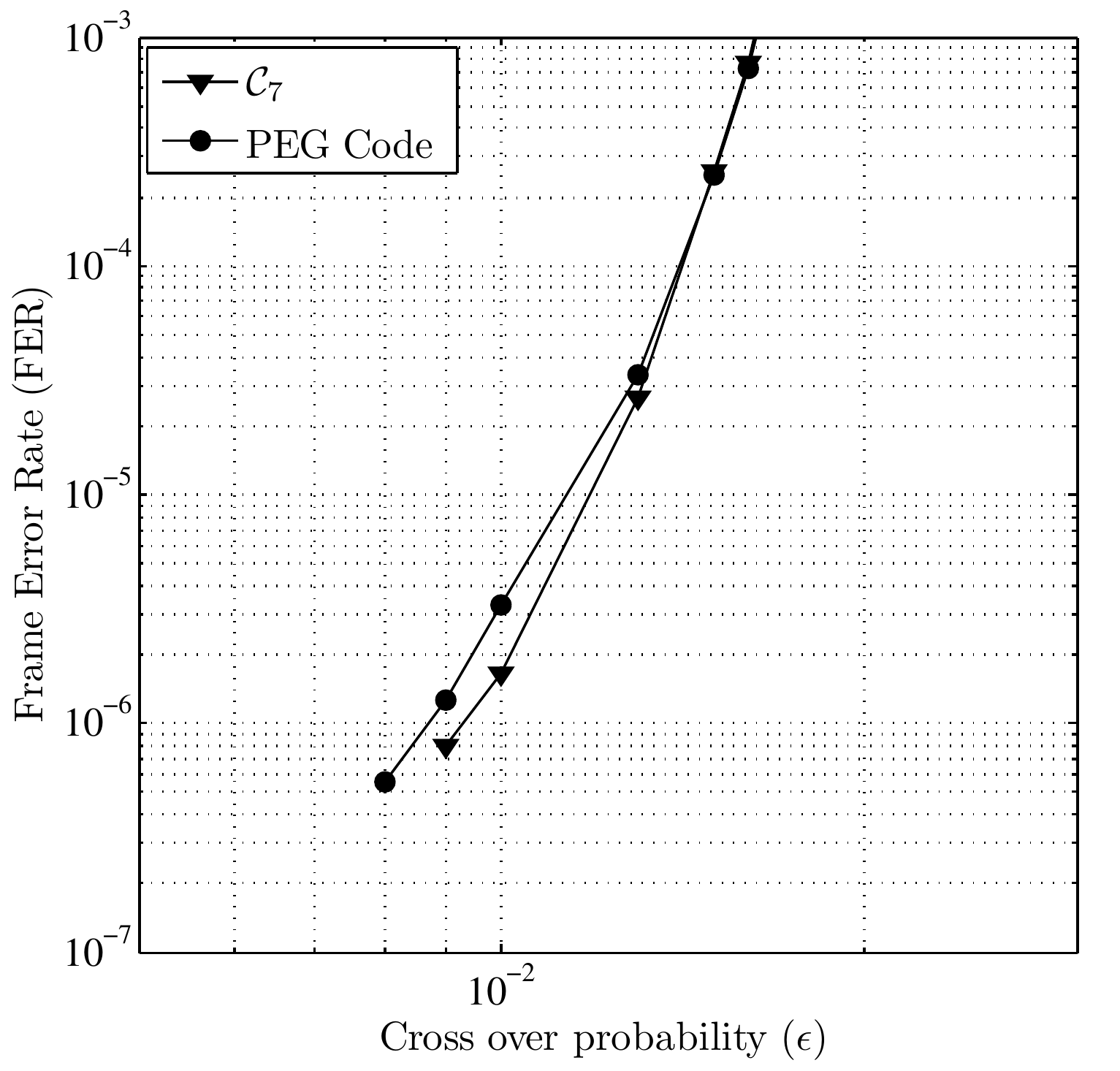}
\caption{Frame error rate performance of codes in Example \ref{ex_no7393} under the SPA on the BSC.}
\label{fig_ne7393fer}
\end{figure}
\end{ex}  

It is worth mentioning that the error rate performance of existing structured regular column-weight-three codes in the literature is at best comparable with the error rate performance of PEG constructed codes. All of our structured codes presented in this paper outperform PEG constructed codes and hence they are candidates for the best known high rate short length regular column-weight-three LDPC codes.

\section{Discussion and Conclusion}\label{discussion}
\begin{figure}
\centering
\includegraphics[width = 3.1 in]{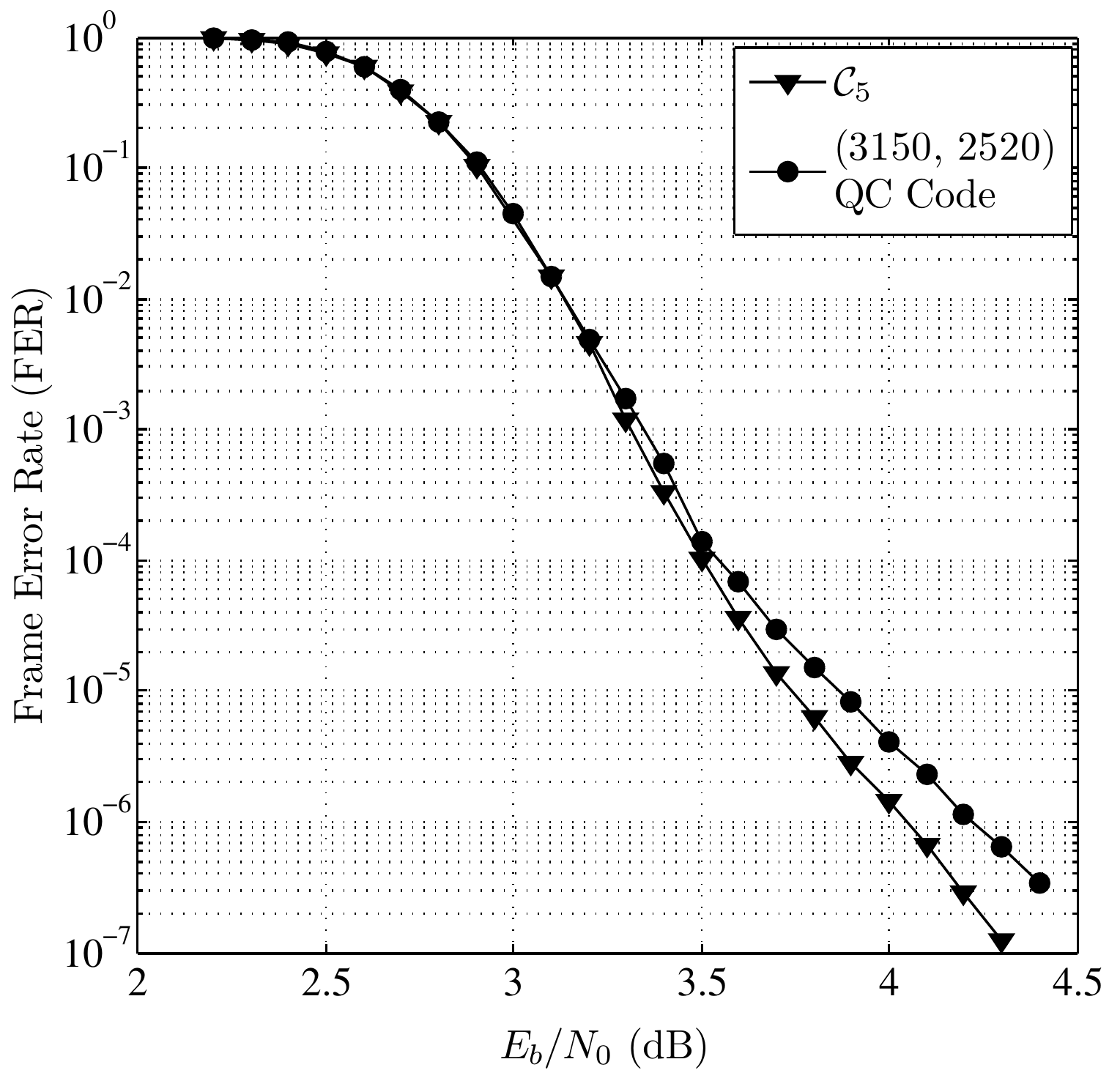}
\caption{Frame error rate performance of codes in Example \ref{ex_no53} under the SPA on the AWGNC.}
\label{fg_awgnNo53}
\end{figure}

Although the codes presented in this paper are optimized for the BSC, they also have excellent performance on the AWGNC. As a demonstration, we show the FER performance of the code $\mathcal{C}_5$ from Example \ref{ex_no53} under the SPA on the AWGNC in  Fig. \ref{fg_awgnNo53}. Recall that the Tanner graph of $\mathcal{C}_5$ has girth $g=8$ and does not contain $(5,3)\{2\}$ and $(8,2)$ trapping sets and that the ($3150, 2520)$ regular QC LDPC code was constructed using array masking proposed in \cite{qcFiniteField_lanLin}. It can be seen that although $\mathcal{C}_5$ was constructed for the BSC, it outperforms the other code, which was constructed for the AWGNC.

We have introduced a new class of structured LDPC codes with a wide range of rates and lengths. More importantly, we have proposed a method to construct codes whose Tanner graphs are free of small trapping sets. These trapping sets are selected based on their relative harmfulness for the decoding algorithms. We have also presented the constructions of regular column-weight-three codes. These codes have excellent performance on both the BSC and the AWGNC, although they were only optimized for the BSC. To the best of our knowledge, these codes outperform the best known short length structured LDPC codes. Our future work includes extending the TSO to include irregular codes and column-weight-four codes as well as the constructions of column-weight-four codes and irregular LDPC codes with low error floor.


\appendices

\section{Implementation of Techniques of Searching for Trapping Sets}\label{sect_search2}
\subsection{Subroutines}
We assume that the following simple subroutines are used in our search algorithms.

\begin{itemize}
\item $\bf Y$ = \textbf{RowIntersectIndex}(${\bf X_1,X_2},\vartheta$).

Let $\bf X_1$ and $\bf X_2$ be matrices. $\bf Y$ is a matrix of two columns. If $(i_1, i_2)$ is a row of $\bf Y$ then the ${i_1}^{\mathrm{th}}$ row of $\bf X_1$ and the ${i_2}^{\mathrm{th}}$ row of $\bf X_2$ share $\vartheta$ common entries.
\item $\bf Y$ = \textbf{OddDegreeChecks}($H,{\bf X}$).

Let $H$ be the parity check matrix corresponding to a Tanner graph $G$ of an LDPC code. Let $\bf X$ be a matrix with each row of $\bf X$ giving a set of variable nodes.  Assume that all the subgraphs induced by variable nodes in rows of $\bf X$ have the same number of odd degree check nodes. $\bf Y$ is a matrix with the same number of rows as $\bf X$. Elements of the ${i}^{\mathrm{th}}$ row of $\bf Y$ are odd degree check nodes in the subgraph induced by the variable nodes in the ${i}^{\mathrm{th}}$ row of $\bf X$. 

\item $\bf Y$ = \textbf{TotalChecksOfDegreeK}($H,{\bf X},\vartheta$).

Let $H$ be the parity check matrix corresponding to a Tanner graph $G$. Let $\bf X$ be a matrix whose elements are variable nodes in $G$. $\bf Y$ is a one-column matrix with the same number of rows as $\bf X$. The element in the ${i}^{\mathrm{th}}$ row of $\bf Y$ is the number of check nodes with degree $\vartheta$ in the subgraph induced by the variable nodes in the ${i}^{\mathrm{th}}$ row of $\bf X$. 

\item $\bf Y$ = \textbf{IsTrappingSet}($H,{\bf X}$).

Let $H$ be the parity check matrix corresponding to a Tanner graph $G$. Let $\bf X$ be a matrix whose elements are variable nodes in $G$. $\bf Y$ is a one-column matrix with the same number of rows as $\bf X$. The element in the ${i}^{\mathrm{th}}$ row of $\bf Y$ is 1 if the variable nodes in the ${i}^{\mathrm{th}}$ row of $\bf X$ form a trapping set and is 0 otherwise.
\end{itemize}  

The above subroutines can be implemented using simple sparse matrix operations and hence are of low complexity.

\subsection{Searching for Cycles of Length $\vartheta$}
Since every trapping set contains at least one cycle, the search for trapping sets always starts with finding cycles in the Tanner graph. All cycles of length $\vartheta$ that contain variable node $v$ can be found by performing the following steps.

\begin{enumerate}
\item Construct the tree of depth $\vartheta/2-1$, taking $v$ as the root using the breadth-first search algorithm \cite{searchBookAlgo}. Let $N_1, N_2,\ldots, N_{d_v}$ be sets of leaf nodes of depth $\vartheta/2-1$ such that all the nodes in $N_i$ are descendants of the $i^\mathrm{th}$ neighbor of $v$. It can be shown that $|N_i|\leq (d_v-1)^{t_1}(d_c-1)^{t_2}$ where
\begin{eqnarray}
t_1 &=& \frac{\vartheta}{4}-\frac{3}{2},t_2 = t_1+1\mathrm{~if~} \vartheta/2 \mathrm{~is~odd}\nonumber\\ 
t_1 &=& t_2 =  \frac{\vartheta}{4}-1\mathrm{~if~} \vartheta/2 \mathrm{~is~even}.\nonumber
\end{eqnarray}
\item For every pair of nodes $o_i\in N_i$, $o_j\in N_j$ and $i\neq j$, determine if they share a common neighbor. If so then a cycle of length $\vartheta$ has been found. If $o_i$ and $o_j$ are check nodes then the cycle is induced by the variable nodes that are ancestors of $o_i$ and $o_j$ and their common neighbor. If $o_i$ and $o_j$ are variable nodes then the cycle is induced by $o_i$ and $o_j$ as well as the variable nodes that are ancestors of $o_i$ and $o_j$. The maximum number of possible pairs $o_i$, $o_j$ is $\sum_{i\neq j}d_v(d_v-1)|N_i||N_j|$.
\end{enumerate}

The two steps described above are executed for every variable node. To further simplify the search, after all the cycles containing $v$ are found, $v$ can be marked so that it is no longer included in Step 1 of the search at other variable nodes. The complexity of searching for cycles is polynomial in the degree of the variable nodes and check nodes but increases only linearly in the code length. Note that our search algorithm not only counts the number of cycles but also records the variable nodes that each cycle contains. For this reason, existing efficient algorithms to count number of cycles in a bipartite graph (for example those proposed in \cite{cycleCoundting_Chugg,cycleMP_Karimi}) can not be applied directly.

\begin{ex}
To illustrate the search algorithm, we list the number of cycles in some popular codes, as well as the run-times of the algorithm on a 2.6 GHz computer in Table \ref{tb_runTmCy}.
\begin{table}[htb]
\caption{Number of Cycles of Several LDPC Codes and Run-time of the Cycle Searching Algorithm on a 2.6 GHz Computer}
\begin{center}
\begin{tabular}{|c||c|c|c|}\hline
Codes					     &Tanner		 &Margulis		&MacKay \\\hline\hline
$n$						  	 &155				 &2168				&4095\\\hline
$(d_v,d_c)$	  		 &(3,6)			 &(3,6)				&(3,17)\\\hline
Number of 6-cycles &0					 &0						&5183\\\hline
Number of 8-cycles &465				 &1320  			&121238\\\hline
Number of 10-cycles&3720			 &11088				&3038421\\\hline
Run-time (Seconds)  &0.007			 &0.23				&28.79\\\hline	
\end{tabular}
\end{center}
\label{tb_runTmCy}
\end{table}
\end{ex}

\subsection{Searching for $(a+1,b-1)$ Trapping Sets Generated by $(a,b)$ Trapping Sets}
Let $\mathcal{T}_1$ be an $(a,b)$ trapping set, $\mathcal{T}_2$ be an $(a+1,b-1)$ trapping set and let $\mathcal{T}_1$ be a parent of $\mathcal{T}_2$. Further, let $\bf T_1$ be a trapping set of type $\mathcal{T}_1$ in the Tanner graph of a code $\mathcal{C}$ and assume that $\bf T_1$ generates a trapping set $\bf T_2$ of type $\mathcal{T}_2$. As discussed in Section \ref{sect_ontology}, $\mathcal{T}_2$ is obtained by adjoining one variable node to $\mathcal{T}_1$. The line-point representation of $\mathcal{T}_2$ is obtained by merging two black shaded nodes in the line-point representation of $\mathcal{T}_1$ with two $\circledast$ nodes in  Fig. \ref{fig_adjGraph}\subref{dl}. Therefore, to search for $\bf T_2$, it is sufficient to search for a variable node that is connected to two odd degree check nodes in the subgraph induced by variable nodes in $\bf T_1$.

Let $\bf X$ be a matrix whose each row contains variable nodes of a $\mathcal{T}_1$ trapping set in the Tanner graph $G$. $H$ is the parity check matrix which defines $\mathcal{C}$. All $\mathcal{T}_2$ trapping sets can be found by performing the following steps.
\begin{enumerate}
\item Find all odd degree check nodes of all $\mathcal{T}_1$ trapping sets: 

$\bf Y_1$ = \textbf{OddDegreeChecks}($H,{\bf X}$).
\item Form $\bf X_1$, a one-column matrix with $n$ rows where the element in the $i^\mathrm{th}$ row is variable node $i$.
\item Form a matrix $\bf Y_2$ whose $i^\mathrm{th}$ row gives all check nodes neighboring to the variable node $i$: 

$\bf Y_2$ = \textbf{OddDegreeChecks}($H,{\bf X_1}$).

\item Find all pairs $(i,j)$ such that the ${i}^{\mathrm{th}}$ row of $\bf Y_1$ and the ${j}^{\mathrm{th}}$ row of $\bf Y_2$ share 2 common entries.

$\bf Y_3$ = \textbf{RowIntersectIndex}(${\bf Y_1,Y_2},2$).
\item If $(i,j)$ is the ${l}^{\mathrm{th}}$ row of $\bf Y_3$, adjoin variable node $j$ to the ${i}^{\mathrm{th}}$ row of ${\bf X}$ to form the ${l}^{\mathrm{th}}$ row of $\bf Y_4$.
\item Determine the number of degree one check nodes in the subgraph induced by variable nodes in each row of $\bf Y_4$ and eliminate the rows of $\bf Y_4$ that do not have $b-1$ degree one check nodes. The matrix $\bf Y$ obtained has each row contain variable nodes that induce a $\mathcal{T}_2$ trapping set in the Tanner graph of the code.

$\bf Y$ = $\bf Y_4$(\textbf{TotalChecksOfDegreeK}($H,{\bf Y_4},1$)$==b-1$).
\end{enumerate}

\subsection{Searching for $(a+2,b)$ Trapping Sets Generated by $(a,b)$ Trapping Sets}
Let $\mathcal{T}_1$ be an $(a,b)$ trapping set, $\mathcal{T}_2$ be an $(a+2,b)$ trapping set and let $\mathcal{T}_1$ be a parent of $\mathcal{T}_2$. Further, let $\bf T_1$ be a trapping set of type $\mathcal{T}_1$ in the Tanner graph of a code $\mathcal{C}$ and assume that $\bf T_1$ generates a trapping set $\bf T_2$ of type $\mathcal{T}_2$. Consider two variable nodes that share a check node. As discussed in Section \ref{sect_ontology}, $\mathcal{T}_2$ is obtained by adjoining these two variable nodes to $\mathcal{T}_1$. The line-point representation of $\mathcal{T}_2$ is obtained by merging two black shaded nodes in the line-point representation of $\mathcal{T}_1$ with two $\circledast$ nodes in  Fig. \ref{fig_adjGraph}\subref{tl}. Therefore, to search for $\bf T_2$, it is sufficient to search for a pair of variable nodes that share a common neighboring check node and each node is connected to one odd degree check node in the subgraph induced by variable nodes in $\bf T_1$.

The search for $(a+2,b)$ trapping sets is very similar to the search for $(a+1,b-1)$ trapping sets described in the previous subsection. In particular, the following modifications should be made:
\begin{itemize}
\item In Step 2, $\bf X_1$ is a two column matrix, each row contains a pair of variable nodes that share a common neighboring check node.
\item In Step 3, the ${i}^{\mathrm{th}}$ row of $\bf Y_2$ gives all degree one check nodes in the subgraph induced by variable nodes in the ${i}^{\mathrm{th}}$ row of $\bf X_1$.
\item In Step 5, variable nodes in the ${j}^{\mathrm{th}}$ row of $\bf X_1$ are adjoined to the ${i}^{\mathrm{th}}$ row of $\bf X$.
\item In Step 6, $b-1$ is replaced by $b$.
\end{itemize}

Since Step 4 does not take into account the case in which two check nodes of a new variable node are merged with two odd degree check node of $\bf T_1$, the subroutine \textbf{IsTrappingSet} is used afterward to eliminate rows of $\bf Y$ that do not contain variable nodes that form a trapping set.

\subsection{Remarks}
The above search procedures may not differentiate among different ($a,b$) trapping sets. For example, all $(5,3)\{1\}$ and $(5,3)\{2\}$ trapping sets are found if they are searched for as trapping sets generated by the $(4,4)$ trapping sets. Similarly, all $(7,3)\{2\}$ and $(7,3)\{3\}$ are found as trapping sets generated by the $(6,4)\{1\}$ trapping sets. If searching for a specific type of trapping set is required, then it is necessary to further analyze the induced subgraph. For example, notice that the $(5,3)\{1\}$ trapping set is a union of a 6-cycle and an 8-cycle, sharing two variable nodes while the $(5,3)\{1\}$ trapping set is a union of two 8-cycle, sharing three variable nodes. Therefore, to search for all $(5,3)\{2\}$ trapping sets from a list of $(4,4)$ trapping sets, one would find all pairs of $(4,4)$ trapping sets that share three variable nodes, using the \textbf{RowIntersectIndex} subroutine. The union of each pair of $(4,4)$ trapping sets is then a set of five variable nodes. Each set of variable nodes forms a $(5,3)\{2\}$ trapping set if its induced subgraph contains three degree one check nodes. Similarly, all $(7,3)\{2\}$ can be found by noticing that they are unions of two $(6,4)\{1\}$ trapping sets, sharing five variable nodes. 
\begin{ex}
We end this section by giving the statistics of small trapping sets present in the random MacKay code of length 4095 along with the running times of the algorithms. These are given in Table \ref{tb_MacKay}. Note that the numbers of six, eight and ten cycles present in the Tanner graph of this code are given in Table \ref{tb_runTmCy}. All the searches were performed in a 2.6 GHz computer.

\begin{table}[htb]
\caption{Number of Small Trapping Sets in MacKay Random Code of Length 4095}
\begin{center}
\begin{tabular}{|c|c|c|}\hline
Trapping Sets	     &Total		 &Run-time (Seconds)\\\hline\hline
$(5,3)\{1\}$  		 &19617		 &0.59      			 \\\hline
$(5,3)\{2\}$  		 &3259		 &12.18      			 \\\hline
$(6,2)\{1\}$  		 &167		   &0.16      			 \\\hline
$(7,1)\{1\}$  		 &2			   &0.05      			 \\\hline
$(6,4)\{1\}$  		 &299636   &55.78      			 \\\hline	
$(7,3)\{2\}$ and $(7,3)\{3\}$ &56309			   &4.21      			 \\\hline
\end{tabular}
\end{center}
\label{tb_MacKay}
\end{table}
\end{ex}

\section{Relations to Array LDPC Codes}\label{sect_arrayCode}
We show that the class of codes described in Section \ref{sec_additive} contains array LDPC codes when the Galois field is a prime field. The parity check matrix of an array LDPC code described by Fan in \cite{arrayCode_fan} is a  $\gamma\times\rho$ subarray of the matrix $\mathcal{H}_{\mbox{arr}}$ of the form
\begin{eqnarray}\label{eq_Harr}
\mathcal{H}_{\mbox{arr}} = \left[
\begin{array}{ccccc}
I&I&I&\cdots&I\\
I&J&J^2&\cdots&J^{q-1}\\
I&J^2&J^4&\cdots&J^{2(q-1)}\\
\vdots\!\!&\vdots&\vdots&\ddots&\vdots\\
I&J^{q-1}&J^{2(q-1)}&\cdots&J^{(q-1)(q-1)}\\
\end{array}
\right],
\end{eqnarray}
where $q$ is an odd prime and $J$ is a $q\times q$ circulant matrix. We now show that $\mathcal{H}_{\mbox{arr}}$ can be obtained by permuting rows and columns of $\mathcal{H}$ in (\ref{eq_Hdef2}), hence array LDPC codes are contained in the new class of LDPC codes proposed in this paper.

Let $q$ be an odd prime. Since the additive group of GF($q$) is cyclic, we can write $\mbox{GF}(q) = \{\beta_{-\infty} = 0,\beta_0 = 1,\beta_1,\ldots,\beta_{q-2}\}$, where $\beta_{i+1}=\beta_i+1$ for $0\leq i\leq q-3$ and $\beta_{-\infty}=\beta_{q-2}+1$. Permute rows and columns of $\mathcal{L}$ to obtain $\mathcal{L}_\beta$, a Latin square that has $(\beta_{-\infty},\beta_0,\beta_1,\ldots,\beta_{q-2})$ as indices of rows from top to bottom and columns from left to right. It can be shown that there exists a permutation matrix $\mathcal{O}$ such that $\mathcal{L}_\beta = \mathcal{O}\mathcal{L}\mathcal{O}$. Replace $\mathcal{L}$ by $\mathcal{L}_\beta$ and let $\mathcal{M}_\beta$ be the sets of images of $\mbox{GF}(q)$ under $f$. It can be seen that:
\begin{itemize}
\item $\mathcal{M_\beta}$ is the set of circulant permutation matrices of size $q \times q$.
\item $f(\beta_{-\infty}) = I$, the $q\times q$ identity matrix.
\item Proposition \ref{prop_addi}, \ref{pmq_prop} and Theorem \ref{te_crossAddi} still holds when $\alpha_t$ is replaced with $\beta_t$ and $P$ is replaced with $P_\beta = \mathcal{O}P\mathcal{O}'$.
\end{itemize}

Finally, permute rows and columns of $\mathcal{W}$ in \ref{eq_Wdef} to obtain $\mathcal{W}_\beta = \mathcal{O}\mathcal{W}\mathcal{O}'$. Then $\mathcal{W}_\beta$ has the form:
\begin{eqnarray}\label{eq_Wbdef}
\mathcal{W}_{\beta} = \left[\begin{array}{ccccc}
0&0&0&\!\cdots&0\\
0&1&\beta_1&\cdots&\beta_{q-2}\\
0&\beta_1&\beta^2_1&\cdots&\beta_1\beta_{q-2}\\
\vdots&\vdots&\vdots&\ddots&\vdots\\
0&\beta_{q-2}&\beta_{q-2}\beta_1&\cdots&\beta^2_{q-2}\\
\end{array}\right], \nonumber
\end{eqnarray}
and hence $f(\mathcal{W}_\beta) = \mathcal{H}_{\mbox{arr}}$, with $J = f(\beta_0) = f(1)$.

\section{Quasi-cyclic LDPC Codes from Cyclic Groups of Permutation Matrices}\label{sect_LinCodes}
In \cite{qcFiniteField_lanLin}, Lan \textit{et. al} give the construction of a class of QC LDPC codes based on the multiplicative groups of Galois fields. We briefly describe this class of codes, but with the formulation introduced in this paper.

Consider the Galois field GF($q$), where $q$ is a power of a prime. Let $\mathcal{L}={[l_{i,j}]}_{i,j\in \mathcal{Q}}$ denote a Latin square defined by the Cayley table of $(\mathcal{Q},\oplus)$ where $\mathcal{Q} = \{1,\alpha,\ldots,\alpha^{q-2}\}$ and $\oplus$ is the multiplicative operation of GF($q$), i.e., $l_{i,j} = i\times j$. Let $\mathcal{M} = \{M_0, M_1, \ldots, M_{q-2}\}$ be the set of images of elements of $\mathcal{Q}$ under $f$. We give the following statements without proofs:
\begin{itemize}
\item $M_0 = I$ is the $(q - 1) \times (q - 1)$ identity matrix.
\item $f(\alpha^{t_1}\alpha^{t_2} ) = f(\alpha^{t_1})f(\alpha^{t_2})$.
\item $\mathcal{M}$ is the cyclic group of permutation matrices of size $(q - 1) \times (q - 1)$ under ordinary matrix multiplication.
\end{itemize}

Define $\mathcal{W}$ and $\mathcal{H}$ as in (\ref{eq_Adef}) and (\ref{eq_Hdef}), where $(\mathcal{Q},\oplus)$ is the multiplicative group of GF($q$). The following theorem gives the necessary and sufficient condition on $\mathcal{\mathcal{W}}$, such that the Tanner graph corresponding to $\mathcal{H}$ has girth at least 6. We omit the proof since it is very similar to the proof of Theorem \ref{te_crossAddi}.
\begin{te}[Cross-multiplication Constraint]\label{te_crossMulti}
The Tanner graph corresponding to $\mathcal{H}$ contains no cycle of length four iff $w_{i_1,j_1}w_{i_2,j_2}\neq w_{i_1,j_2}w_{i_2,j_1}$ for any $1\leq i_1,i_2 \leq \mu$; $1\leq j_1,j_2 \leq \eta$; $i_1\neq i_2$; $j_1\neq j_2$.
\end{te}

For any pair ($\gamma,\rho$) of positive integers with $1\leq \gamma,\rho \leq q$, let $H$ be a $\gamma\times\rho$ subarray of $\mathcal{H}$. Then $H$ is a $\gamma q\times \rho q$ matrix over GF(2) which is also free of cycles of length 4. $H$ has constant column weight $d_v = \gamma$ and row weight $d_c = \rho$. The null space of $H$ gives a regular structured LDPC code $\mathcal{C}$ of length $\rho q$ with rate at least $R = (\rho-\gamma)/\rho$ \cite{ldpcBook_gallager}.

\textit{Remarks:} We can adjoint the zero element of GF($q$) to the set $\mathcal{Q}$ to obtain $\mathcal{Q}' = \mathcal{Q}\cap\{0\}$ and define $f(0) = Z$, the all zero matrix of size $(q - 1) \times (q - 1)$. Theorem \ref{te_crossMulti} still holds for $\mathcal{W}$ defined on $\mathcal{Q}'$. In this case, the cross-multiplication constraint  is equivalent to the \textit{$\alpha$-multiplied row constraints} given in \cite{qcFiniteField_lanLin}. It is almost obvious that Latin squares obtained from the Cayley table of the additive group of GF($q$) satisfy the cross-multiplication constraint. This concept is used in \cite{codesLatinSquare_abdel} to obtain a class of QC LDPC codes on Latin squares. 

\section{Minimum Distance of Code Constructed from GF($2^\vartheta$)}
The structured LDPC codes proposed in this paper include codes of length $2^\vartheta$, which allow hardware implementation to be further simplified. Unfortunately, the minimum distances of these codes are upper bounded by 8. 
\begin{te}\label{te_mind2pm}
Let $\mathcal{C}$ be an LDPC code defined in Section \ref{sec_additive} with $\gamma=3$. If $q = 2^\vartheta$ where $\vartheta \in \mathbb{N}, \vartheta>1$ then the minimum distance of $\mathcal{C}$ is at most 8.
\end{te}
\IEEEproof
Let $H$ be the parity check matrix of $\mathcal{C}$, where $\mathcal{C}$ is an LDPC code defined in Section \ref{sec_additive} with $q = 2^\vartheta$. We know that $H = f(\mathcal{W})$, where $\mathcal{W}$ is a matrix over GF($q$). Let $W$ be a matrix formed by any two columns of $W$. WLOG assume that
\begin{eqnarray}
W = \left[\begin{array}{cc}
\alpha^{t_1}&\alpha^{t_4}\\
\alpha^{t_2}&\alpha^{t_5}\\
\alpha^{t_3}&\alpha^{t_6}
\end{array}\right].\nonumber
\end{eqnarray}
Let $i_1 = \xi \in \mathrm{GF}(q)$ and let
\begin{eqnarray}
\begin{array}{cc}
j_1 = \xi + \alpha^{t_1},& j_2 = \xi + \alpha^{t_2} + \alpha^{t_4} + \alpha^{t_5},\\ 
j_5 = \xi + \alpha^{t_2},& j_3 = \xi + \alpha^{t_3} + \alpha^{t_4} + \alpha^{t_6},\\
j_9 = \xi + \alpha^{t_3},& j_6 = \xi + \alpha^{t_1} + \alpha^{t_4} + \alpha^{t_5},\\
i_5 = \xi + \alpha^{t_1} + \alpha^{t_4}, & j_8 = \xi + \alpha^{t_3} + \alpha^{t_5} + \alpha^{t_6},\\
i_6 = \xi + \alpha^{t_2} + \alpha^{t_5}, & j_{11} = \xi + \alpha^{t_1} + \alpha^{t_4} + \alpha^{t_6},\\
i_7 = \xi + \alpha^{t_3} + \alpha^{t_6}, & j_{12} = \xi + \alpha^{t_2} + \alpha^{t_5} + \alpha^{t_6},
\end{array}\nonumber
\end{eqnarray}

\begin{eqnarray}
i_2 &=& \xi + \alpha^{t_1} + \alpha^{t_2} + \alpha^{t_4} + \alpha^{t_5}, \nonumber\\
i_3 &=& \xi + \alpha^{t_1} + \alpha^{t_3} + \alpha^{t_4} + \alpha^{t_6}, \nonumber\\
i_4 &=& \xi + \alpha^{t_2} + \alpha^{t_3} + \alpha^{t_5} + \alpha^{t_6}, \nonumber\\
j_7 &=& \xi + \alpha^{t_1} + \alpha^{t_2} + \alpha^{t_3} + \alpha^{t_4} + \alpha^{t_6},\nonumber\\
j_{10} &=& \xi + \alpha^{t_1} + \alpha^{t_2} + \alpha^{t_3} + \alpha^{t_4} + \alpha^{t_5},\nonumber\\
i_8 &=& \xi + \alpha^{t_1} + \alpha^{t_2} + \alpha^{t_3} + \alpha^{t_4} + \alpha^{t_5} + \alpha^{t_6}.\nonumber
\end{eqnarray}
Then the following equations hold since $\alpha^t + \alpha^t = 0 \;\forall \alpha^t \in \mathrm{GF}(q)$ 
\begin{eqnarray}
\begin{array}{ccc}
i_1-j_1 = \alpha^{t_1}, & i_1-j_5 = \alpha^{t_2}, & i_1-j_9 = \alpha^{t_3},\nonumber\\
i_2-j_2 = \alpha^{t_1}, & i_2-j_6 = \alpha^{t_2}, & i_2-j_{10} = \alpha^{t_3},\nonumber\\
i_3-j_3 = \alpha^{t_1}, & i_3-j_7 = \alpha^{t_2}, & i_3-j_{11} = \alpha^{t_3},\nonumber\\
i_4-j_4 = \alpha^{t_1}, & i_4-j_8 = \alpha^{t_2}, & i_4-j_{12} = \alpha^{t_3},\nonumber\\
i_5-j_1 = \alpha^{t_4}, & i_6-j_5 = \alpha^{t_5},	& i_7-j_9 = \alpha^{t_6},\nonumber\\
i_5-j_6 = \alpha^{t_5}, & i_6-j_{12} = \alpha^{t_6}, & i_8-j_4 = \alpha^{t_4},\nonumber\\
i_5-j_{11} = \alpha^{t_6}, & i_7-j_3 = \alpha^{t_4}, & i_8-j_7 = \alpha^{t_5},\nonumber\\
i_6-j_2 = \alpha^{t_4} & i_7-j_8 = \alpha^{t_5} &i_8-j_{10} = \alpha^{t_6}.\nonumber
\end{array}
\end{eqnarray}

Recall that $f(\alpha^t)$ is a permutation matrix whose rows and columns are indexed using elements of GF($q$), and that its entries $m^{(t)}_{i,j}=1$ if and only if $i-j = \alpha^t$. Since the above equations hold, there exist eight variable nodes in the Tanner graph corresponding to $f(W)$ with line-point representation shown in  Fig. \ref{fig_cwg8}\subref{w8_1}. In other words, the Tanner graph of $\mathcal{C}$ contains the $(8,0)\{1\}$ weight-eight codewords.
\endIEEEproof

Clearly, the Tanner graph of $\mathcal{C}$ contains trapping sets that are parents of the $(8,0)\{1\}$ codeword, i.e., it contains the $(7,3)\{2\}$ and $(6,4)\{1\}$ trapping sets shown in  Fig. \ref{fig_64andChild}\subref{641_lp} and \ref{fig_64andChild}\subref{732_lp} and eight cycles. Consequently, the Tanner graph of $\mathcal{C}$ has girth at most 8. From the proof of Theorem \ref{te_mind2pm}, it can be seen that the number of $(8,0)\{1\}$ codewords of $\mathcal{C}$ is lower bounded by ${\binom{\rho}{2}}2^{\vartheta-2}$, where $\rho$ is the row weight of $\mathcal{C}$. We also note that for all the column-weight-three codes of girth $g=8$ that we have constructed from GF($2^\vartheta$), this lower bound gives the exact number of $(8,0)\{1\}$ codewords. Moreover, we found no $(8,0)\{2\}$ codewords (the line-point representation of the $(8,0)\{2\}$ codeword is shown in  Fig. \ref{fig_cwg8}\subref{w8_2}) in these codes. Therefore, the total number of weight-eight codewords in these codes is ${\binom{\rho}{2}}2^{\vartheta-2}$.

\bibliographystyle{IEEEtran}
\bibliography{reference}


\end{document}